\documentclass[aps,reprint,superscriptaddress,longbibliography]{revtex4-1}

\usepackage{hyperref}
\usepackage{outlines}
\usepackage{fancyvrb}
\usepackage{amsmath}
\usepackage{hyperref}
\usepackage{braket}
\usepackage{bbold}
\usepackage{amssymb}
\usepackage{graphicx}
\usepackage[usenames, dvipsnames]{color}
\graphicspath{{/Users/hlerkristjansson/Desktop/qonn_paper_hler_26aug/}{C:\Users}}

\begin{document}


\title{Quantum generalisation of feedforward neural networks}


\author{Kwok Ho Wan}
\affiliation{Blackett Laboratory,
  Imperial College London, London, SW7 2AZ, United Kingdom}
\affiliation{London Institute for Mathematical Sciences, 35a South Street
Mayfair, London, W1K 2XF, United Kingdom}
\author{Oscar Dahlsten}
\affiliation{Blackett Laboratory,
  Imperial College London, London, SW7 2AZ, United Kingdom}
\affiliation{Clarendon Laboratory,
  University of Oxford, Parks Road, Oxford, OX1 3PU, United Kingdom}
\affiliation{London Institute for Mathematical Sciences, 35a South Street
Mayfair, London, W1K 2XF, United Kingdom}
\author{Hl\'er Kristj\'ansson}
\affiliation{Blackett Laboratory,
  Imperial College London, London, SW7 2AZ, United Kingdom}
\affiliation{London Institute for Mathematical Sciences, 35a South Street
Mayfair, London, W1K 2XF, United Kingdom}
\author{Robert Gardner}
\affiliation{Blackett Laboratory,
  Imperial College London, London, SW7 2AZ, United Kingdom}
\affiliation{London Institute for Mathematical Sciences, 35a South Street
Mayfair, London, W1K 2XF, United Kingdom}
\author{M.S. Kim}
\affiliation{Blackett Laboratory,
  Imperial College London, London, SW7 2AZ, United Kingdom}


\date{\today}

\begin{abstract}
We propose a quantum generalisation of a classical neural network. The classical neurons are firstly rendered reversible by adding ancillary bits.
Then they are generalised to being quantum reversible, i.e.\ unitary. (The classical networks we generalise are called feedforward, and have step-function activation functions.)
The quantum network can be trained efficiently using gradient descent on a cost function to perform quantum generalisations of classical tasks. We demonstrate numerically that it can: (i) compress quantum states onto a minimal number of qubits, creating a quantum autoencoder, and (ii) discover quantum communication protocols such as teleportation. Our general recipe is theoretical and implementation-independent. The quantum neuron module can naturally be implemented photonically.
\end{abstract}

\pacs{}

\maketitle


\section*{Introduction}
Artificial neural networks mimic biological neural networks to perform information processing tasks. 
 They are highly versatile, applying to vehicle control, trajectory
prediction, game-playing, decision making, pattern recognition (such
as facial recognition, spam filters), financial time series
prediction, automated trading systems, mimicking unpredictable
processes, and data mining~\cite{Nielsen15, Azoff94}. 
The networks can be trained to perform tasks without the programmer necessarily 
detailing how to do it. Novel techniques for training networks of many layers (deep networks) is credited with giving impetus to the neural networks approach~\cite{LecunBengioHinton15}. 

The field of quantum machine learning is rapidly developing though the focus has aruably not been on the connection to neural networks. Quantum machine learning, see e.g.~\cite{LloydMR13, LloydMR13ii, Montanaro15, Aaronson15, GarneroneZL12, HarrowHL09, LloydGZ16, RebenstrostML13, WiebeBL12,Adcock15, HeimRIT15, GrossYFBE10, Dunjko16, Wittek14} employs quantum information processing (QIP)~\cite{NielsenChuang00}. QIP uses quantum superpositions of states with the aim of faster processing of classical data as well as tractable simulation of quantum systems.  In a superposition each bit string is associated with two numbers: the probability of the string and the {\em phase}~\cite{GarnerDNMV15}, respectively. The phase impacts the future probabilities via a time evolution law. There are certain promising results that concern quantum versions of recurrent neural networks, wherein neurons talk to each other in all directions rather than feeding signals forward to the next layer, e.g. with the purpose of implementing quantum simulated annealing~\cite{Lechnere1500838, quantum-deep-learning, HeimRIT15, GarneroneZL12}.  In~\cite{Schuld14} several papers proposing quantum neural network designs are discussed and critically reviewed.  A key challenge to overcome is the clash between the nonlinear, dissipative dynamics of neural network computing and the linear, reversible dynamics of quantum computing~\cite{Schuld14}. A key reason for wanting well-functioning quantum neural networks is that these could do for quantum inputs what classical networks can do for classical inputs, e.g. compressing data encoded in quantum superpositions to a minimal number of qubits.  

We here accordingly focus on creating quantum generalisations of
classical neural networks, which can take quantum inputs and process them
coherently. Our networks contribute to a research direction known as {\em
  quantum learning} ~\cite{Bisio10, Sasaki02, Sentis15, Banchi16, Palittapongarnpim16} which concerns learning and optimising with truly quantum objects. The networks provide a route to harnessing the powerful neural network paradigm for this purpose. Moreover they are strict generalisations of the classical networks, providing a clear framework for comparing the power of quantum and classical neural networks. 

The networks generalise classical neural networks to the quantum case in a similar sense to how quantum computing generalises classical computing.  We start with a common classical neural network family: feedforward perceptron networks. We make the invidual neurons reversible and then naturally generalise them to being quantum reversible (unitary). This resolves the classical-quantum clash mentioned above from~\cite{Schuld14}. An efficient training method is identified: global gradient descent for a quantum generalisation of the cost function, a function evaluating how close the outputs are to the desired outputs. To illustrate the ability of the quantum network we apply it to (i) compressing information encoded in superpositions onto fewer qubits (an autoencoder) and (ii) re-discovering the quantum teleportation protocol---this illustrates that the network can work out QIP protocols given only the task. To make the connection to physics clear we describe how to simulate and train the network with quantum photonics. 

We proceed as follows. Firstly, we describe the recipe for generalising the classical neural network. Then it is demonstrated how the network can be applied to the tasks mentioned above, followed by a design of a quantum photonic realisation of a neural module. We discuss the results, followed finally by a summary and outlook.  

\section*{Quantum neural networks}
Classical neural networks are composed of elementary units called neurons. We begin with describing these, before 
detailing how to generalise them to quantum neurons.  

\subsection*{The classical neuron}
\begin{figure}[h]
\includegraphics[height=2cm]{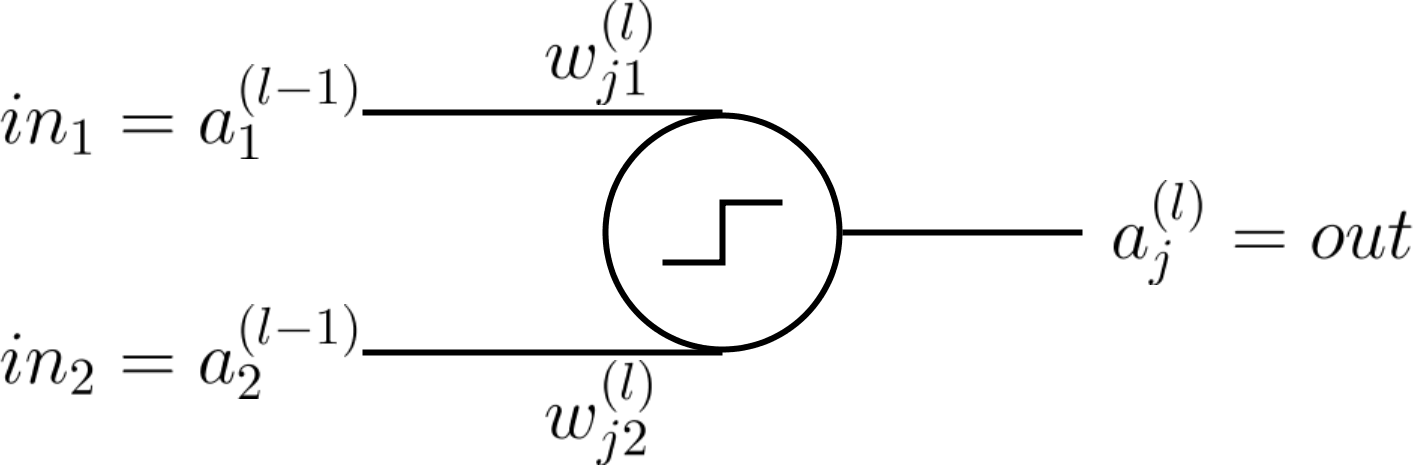}
\caption{\label{fig:classical_neuron}A classical neuron taking two
  inputs $in_{1}$ and $in_{2}$ and giving a corresponding output
  $out$~\cite{Nielsen15}. $a_j^{(l)}$ labels the output of the $j^{th}$ neuron in the $l^{th}$ layer of the network. }
\end{figure}

\noindent A classical neuron is depicted in FIG.~\ref{fig:classical_neuron}.
In this case, it has two inputs (though there could be more). There is one output, which depends on the 
inputs (bits in our case) and a set of weights (real numbers): if the weighted sum of inputs is above a set threshold, the output is 1, else it is 0. 

We will use the following standard general notation. The $j^{th}$ neuron in the
$l^{th}$ layer of a network takes a number of inputs, $a^{(l-1)}_{k}$, where $k$ labels the input.
The inputs are each multiplied by a corresponding weight, $w^{(l)}_{jk}$, and an output, $a_{j}^{(l)}$, is fired as a function of the weighted
input $z_{j}^{(l)}=\sum_{k=1}^{n} w^{(l)}_{jk} a^{(l-1)}_{k} $, where $n$ is the number of inputs to the neuron (FIG.~\ref{fig:classical_neuron}). The function relating the output to
the weighted input is called the activation function, which has most
commonly been a Heaviside step function or a sigmoid
~\cite{Nielsen15}. For example, the neuron in FIG.~\ref{fig:classical_neuron} with a Heaviside activation
function gives an output of the form:
\begin{equation}
a_{j}^{(l)} = \left\{\begin{matrix}
1, ~$if$~ z_{j}^{(l)} > 0.5\\ 
\!\!\! 0, ~$otherwise.$\\ 
\end{matrix}\right.
\end{equation}
This paper aims to generalise the classical neuron to a quantum mechanical
one. In the absence of measurement, quantum mechanical processes are
required to be reversible, and more specifically, unitary, in a closed
quantum system~\cite{Feynman86, NielsenChuang00}. This
suggests the following procedure for generalising the neuron first to
a reversible gate and finally to a unitary gate: 

{\em Irreversible $\rightarrow$ reversible:} For an $n$-input classical neuron having $(in_1, in_2, ..., in_n )\rightarrow out$,
 create a classical reversible gate taking ($in_1, in_2, ..., in_n, 0) \rightarrow (in_1, in_2, ..., in_n, out)$. Such an operation can
  always be represented by a permutation matrix~\cite{Muthukrishnan99}. This is a clean way of rendering
the classical neuron reversible. The extra `dummy' input bit is used to make it reversible~\cite{Feynman86}; in particular, some of the `2 bits in -- 1 bit out' functions the neuron can implement require 3 bits to be made reversible in this manner.

{\em Reversible $\rightarrow$ unitary:} Generalise the classical reversible gate to a quantum unitary taking input ($\ket{\psi_{in}}_{1,2,...,n}\ket{0}) \rightarrow  \ket{\psi_{out}}_{1,2,...,n,out}$, such that the final output qubit is the output of interest. This is the natural way of making a permutation matrix unitary. 

\begin{figure}[h]
\center
\includegraphics[width=\linewidth]{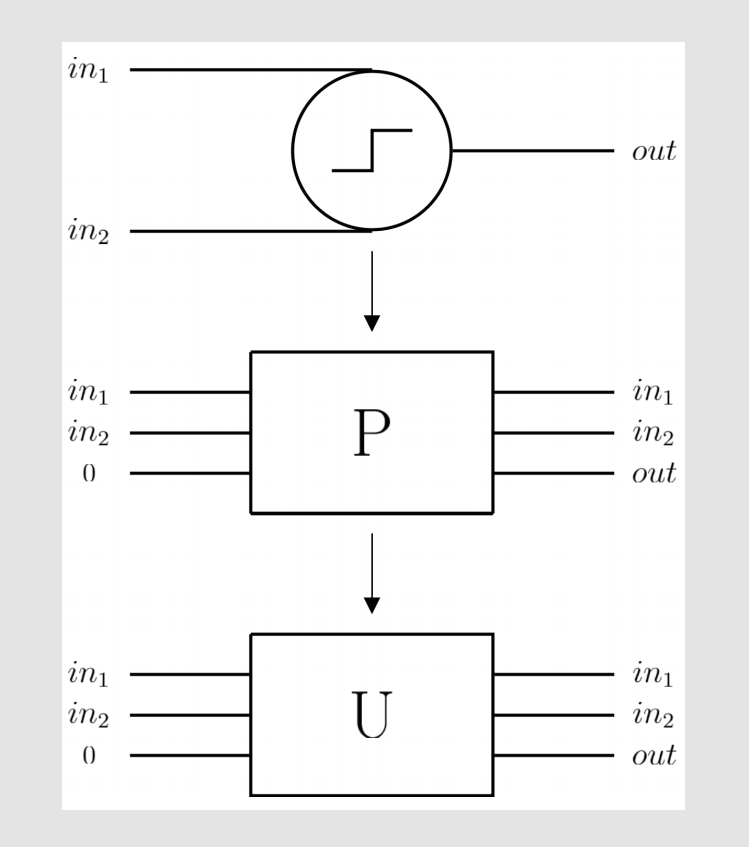}
\caption{\label{fig:neuron_generalisation}Diagram summarising our method
  of generalising the classical irreversible neuron with Heaviside
  activation function, first to a reversible neuron represented by a
  permutation matrix (P), and finally to a quantum reversible
  computation, represented by a unitary operator (U).}
\end{figure}

If the input is a mixture of states in the computational basis and the unitary a permutation matrix~\cite{CurtisReiner62}, the output qubit will be a mixture of $\ket{0}$ or $\ket{1}$: this we call the {\em classical special case}. This way the quantum neuron can simulate any classical neuron as
defined above. The generalisation recipe summarised in
FIG.~\ref{fig:neuron_generalisation} also illustrates how any irreversible classical computation can be recovered as a special case from reversible classical computation (by ignoring the dummy and
copied bits), which in turn can be recovered as a special case from quantum computation.
\subsubsection*{The network}

\begin{figure}[h]
\includegraphics[height=3cm]{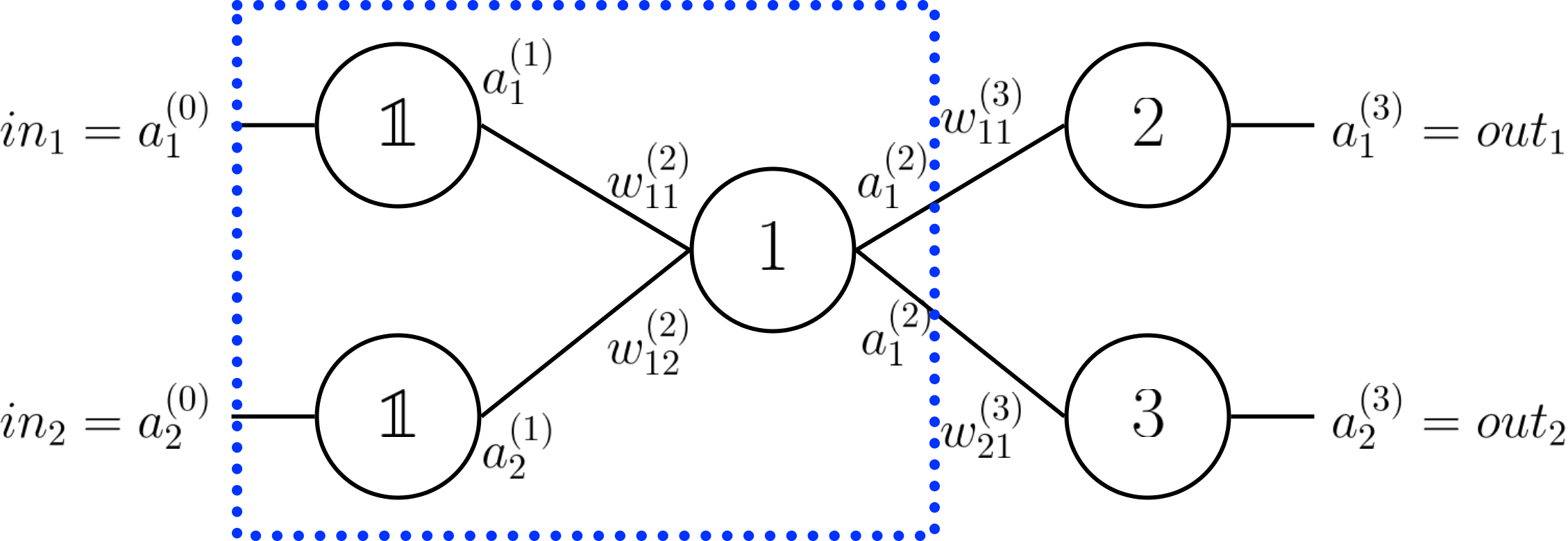}
\caption{\label{fig:cl_autoencoder}A classical autoencoder taking two
  inputs $in_{1}=a^{(0)}_{1}$ and $in_{2}=a^{(0)}_{2}$ and compressing
    them to one hidden layer output $a^{(l)}_{1}$. The final output layer
   is used in training and is trained to reconstruct the inputs.
The notation here is
  in accordance with \cite{Nielsen15}. The blue box represents the data
  compression device after the training procedure.}
\end{figure}

In order to form a neural network, classical neurons are connected
together in various configurations. Here, we consider feedforward
classical networks, where neurons are arranged in layers and each
neuron in the $l^{th}$ layer is connected to every neuron in the
$(l-1)^{th}$ and $(l+1)^{th}$ layers, but with no connections within
the same layer. For an example of such a classical network, see
FIG.~\ref{fig:cl_autoencoder}. Note that in this case the same output of a single neuron is sent
to all the neurons in the next layer~\cite{Nielsen15, Azoff94}. 

To make the copying reversible, in line with our approach of firstly making the classical neural network reversible, we propose the recipe:
\\\\
{\em Irreversible $\rightarrow$ reversible:} For a classical irreversible copying operation of a bit $b \rightarrow (b, b)$,
create a classical reversible gate, which can be represented by a
 permutation matrix~\cite{Feynman86}, taking $(b, 0) \rightarrow (b, b)$.
\\\\
In the quantum case the no-cloning theorem shows one cannot do this in the most naive way
\cite{NielsenChuang00}. For a 2-qubit case, one can use a CNOT for
example to copy in the classical computational basis~\cite{Feynman86}: $\ket{b}\ket{0}\rightarrow \ket{b}\ket{b}$, if $\ket{b}\in \{\ket{0}, \ket{1} \}$. Thus one may consider replacing the copying with a CNOT. However when investigating applications of the network we realised that there are scenarios (the autoencoder in particular) where entanglement between different neurons is needed to perform the task. We have therefore chosen the following definition:
\\\\
{\em Reversible $\rightarrow$ unitary:} The classical CNOT is
generalised to a general 2-qubit `fan-out' unitary $U_F$, with one dummy input set
to $\ket{0}$, such that $\ket{b}\ket{0}\rightarrow U_F\ket{b}\ket{0}$. As this unitary does not in general copy quantum states that are non-orthogonal we call it a `fan-out' operation rather than a copying operation, as it 
distributes information about the input state into several output qubits. Note that a quantum network would be \emph{trained} to choose the unitary in question.

\subsubsection*{Efficient training with gradient descent}

A classical neural network is trained to perform particular tasks. This
is done by randomly initialising the weights and then propagating
inputs through the network many times, altering the weights after each
propagation in such a way as to make the network output closer to the
desired output. A cost function, $C$, relating the network output to the
desired output is defined by
\begin{equation}
\label{eq:cost}
C = \frac{1}{2}\Big|\vec{y}^{\:(L)}-\vec{a}^{\:(L)}\Big|^2,
\end{equation}
where $\vec{y}^{\:(L)}$ is a vector of the desired outputs from each of
the final layer $l=L$ neurons and $\vec{a}^{\:(L)}$ is the vector of actual
outputs, which depends on the network weights, and $\Big|(.)\Big|$ is the $l^2$-norm. The cost function is
minimised to zero when the weights propagate the input in such a way
that the network output vector equals the desired output
vector. 

Since the weights are continuous variables, the numerical partial derivatives of
the cost function w.r.t. each weight can be found by approximating $\frac{\partial C}{\partial w}\approx \frac{ C(w+\epsilon)-C(w)}{ \epsilon}$. After each
propagation, these partial derivatives are computed and the weights
are altered in the direction of greatest decrease of the cost
function. Specifically, each weight $w^{(l)}_{jk}$ is increased by
$\mathrm{\delta} w^{(l)}_{jk}$, with
\begin{equation}
\label{eq:learningrule}
\mathrm{\delta} w^{(l)}_{jk}=-\eta\frac{\partial C}{\partial w^{(l)}_{jk}},
\end{equation}
where $\eta$ is an adjustable non-negative parameter. This training
procedure is known as gradient descent \cite{Nielsen15}.

Note that gradient descent normally also requires a continuous and differentiable activation
function, to allow small changes in the weights to relate to small
changes in the cost. For this reason, the Heaviside activation function has traditionally been replaced
by a sigmoid function \cite{Nielsen15, Azoff94}. Nevertheless, gradient descent
has also been
achieved using Heaviside activation functions, by taking the weights
as Gaussian variables and taking partial derivatives w.r.t. the means
and standard deviations of the appropriate Gaussian distributions
\cite{BartlettDowns92, DownsGaynier95}. 

In the reversible generalisation, where each neuron is replaced by a
permutation matrix, we find that the output is no longer a function of
the inputs and \emph{continuous} weights, but rather of the inputs and a
\emph{discrete set} of permutation matrices. However, in the
generalisation to unitaries, for a gate with $n$ inputs and outputs,
there exist an infinite number of unitaries, in contrast with the discrete
set of permutation matrices. This means that the unitaries can be
parametrised by continuous variables, which once again allows the application of 
gradient descent.

Given that any unitary matrix $U$ can be expressed as $U=e^{iH}$, where $H$
is a Hermitian matrix~\cite{NielsenChuang00}, and that such matrices can be written as linear 
combinations of tensor products of the Pauli matrices and the identity, it follows that a general $N$-qubit unitary can be expressed as
\begin{equation}
\label{eq:u_n}
U_{N} \! = \! \exp \! \Bigg[ \! i \Bigg(\!
\sum_{j_{1},...,j_{N}=0,...,0}^{3,...,3} \! \! \! \! \! \! \! \! \!
\alpha_{j_{1},...,j_{N}} \! \times \! (\sigma_{j_{1}} \!
\otimes ... \otimes \sigma_{j_{N}}) \! \Bigg) \! \Bigg],
\end{equation}
where $\sigma_{i}$ are the Pauli matrices for $i \in \{1,2,3\}$ and
$\sigma_0$ is the $2\times 2$ identity matrix.
This parametrisation allows the use of the training rule of Eq.~\ref{eq:learningrule}, but replacing the weight $w^{(l)}_{jk}$ with a general parameter $\alpha_{j_{1},...,j_{N}}$ of the unitary $U_{N}$:
\begin{equation}
\label{eq:learningruleq}
\mathrm{\delta} \alpha_{j_{1},...,j_{N}}=-\eta\frac{\partial C}{\partial \alpha_{j_{1},...,j_{N}}}.
\end{equation}
 
A simpler and less general form of $U_{N}$ has been sufficient for the tasks discussed in this paper: 
\begin{equation}
\label{eq:u_nrestr}
U_{3} = \sum_{j=1}^4\ket{\tau_j}{\bra{\tau_j}\otimes T_j},
\end{equation}
where $\{\ket{\tau_j}\}_{j=1}^4= \{ V\ket{00},V\ket{01}, V\ket{10}, V\ket{11}\}$. $V$ is a general 2-qubit unitary of the form of Eq.~\ref{eq:u_n}. Each $T_j$ is similarly a general 1-qubit unitary and one can see, using the methods of~\cite{Rowell04} on Eq.~\ref{eq:u_n}, that this can be expressed as a linear combination of the
Pauli matrices, $\sigma_{j}$:
\begin{equation}
U_{1-qubit}=e^{i \alpha_{0}} \Bigg( \cos
\Omega ~\mathbb{1} + i
\frac{\sin{\Omega}}{\Omega} \sum_{j=1}^{3} \alpha_{j} \sigma_{j} \Bigg),
\end{equation}
where $\Omega = \sqrt{\alpha_{1}^2+\alpha_{2}^2+\alpha_{3}^2}$~\cite{Rowell04}. 
To extend this to higher dimensional unitaries, see e.g.~\cite{Hedemann13}.

The cost function we use for the quantum neural networks is, with experimental feasibility in mind, determined by the expectation values of local Pauli matrices $ (\sigma_{1}, \sigma_{2}, \sigma_{3})$ on individual output qubits, $j$. It has the form 
\begin{equation}
C=\sum_{i,j} f_{ij} (\langle \sigma_i^{(j)} \rangle_{\mathrm{actual}} -\langle \sigma_i^{(j)} \rangle_{\mathrm{desired}})^2
\end{equation}
where $f_{ij}$ is a real non-negative number (in the examples to follow $f_{ij}\in \{0,1\}$). We note in the classical mode of operation, where the total density matrix state is diagonal in the computational basis, only $\sigma_3$ will have non-zero expectation, and the cost function becomes the same as in the classical case (Eq. \ref{eq:cost}) up to a simple transformation.

It is important to note that the number of weights grow polynomially in the number of neurons. 
 Each weight shift is determined by evaluating the cost function twice to get the RHS of Eq.~\ref{eq:learningruleq}. Thus the number of evaluations of the cost function for a given iteration of the gradient descent grows polynomially in the number of neurons. The training procedure is efficient in this sense. We do not here attempt to provide a proof that the convergence to zero cost-function, where possible, will always take a number of iterations that grows polynomially in the number of neurons.  Note also that the statements about the efficiency of the training procedure refer to the physical implementation with quantum technology: the simulation of quantum systems with a classical computer is, with the best known methods, in general inefficient. 

\subsection*{Example: Autoencoder for data compression}
We now demonstrate applications of our quantum generalisation of
neural networks described in the previous section. We begin with autoencoders. These compress an input signal from a given set of possible inputs onto a smaller number of bits, and are `work-horses' of classical machine learning~\cite{Azoff94}. 

\subsubsection*{Classical autoencoder}
 Autoencoders are commonly achieved by
a feedforward neural network with a bottleneck in the form of a layer with fewer neurons than the input layer. The network is trained to recreate the signal at a later layer, which necessitates reversibly compressing it (as well as possible) to a bit size equal to the number of neurons in the bottleneck layer~\cite{Azoff94}. The bottleneck layer size can be varied as part of the training to find the smallest compression size possible, which depends on the data set in question. After the training is complete, the post-bottleneck part of the network can be discarded and the compressed output taken directly from after the bottleneck. 

In FIG.~\ref{fig:cl_autoencoder} a basic autoencoder
designed to compress two bits into a single bit is shown. (Here the number of input bits, $j_{max}=2$.) The basic training procedure consists of creating a cost function:
\begin{equation}
C = \sum_{j = 1}^{j_{max}} (in_{j}-out_{j})^{2},
\end{equation}
with which the network is trained using the learning rule of Eq.~\ref{eq:learningrule}. If the outputs are identical to the inputs (to within
numerical precision), the network is fully
trained. The final layer is then removed, revealing the
second last layer, which should enclose the compressed data. The number of neurons in a given hidden layer for a classical
neuron will not exceed $j_{max}$. Once the network is trained, the removal of the
post-bottleneck layer(s) will yield a second last layer of fewer neurons, achieving dimensional reduction~\cite{Azoff94}.

\subsubsection*{Quantum autoencoder}
\begin{figure}[h]
\includegraphics[height=3.2cm]{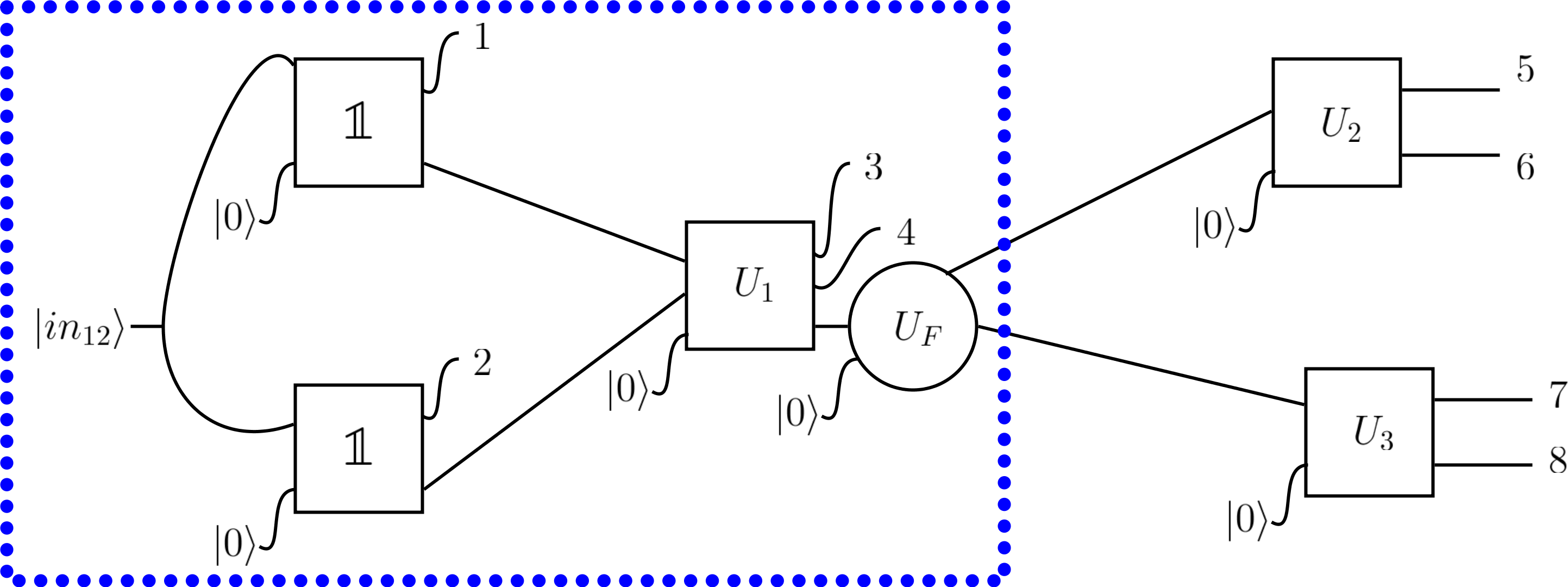}
\caption{\label{fig:quantum_autoencoder}Neural network implementing a quantum
  autoencoder that can accomodate two input qubits that are
  entangled. The blue box represents the quantum compression device after training.}
\end{figure}

We now generalise the classical autoencoder as shown in
FIG.~\ref{fig:cl_autoencoder} to the quantum case. We generalise the neurons
labelled 1, 2 and 3 in FIG.~\ref{fig:cl_autoencoder} into unitary matrices $U_{1}$, $U_{2}$
and $U_{3}$, respectively, with the addition of a `fan-out' gate,
$U_{F}$, as motivated in the previous sections. The result is shown in FIG.~\ref{fig:quantum_autoencoder} as a quantum circuit
model. (We follow the classical convention that this neural
network is drawn with the input neurons as well, but they are identity operators
which let the inputs through regardless, and can be ignored in the
simulation of the network.) The input state of interest $\ket{in_{12}}$ is on 2 qubits, each fed into a different neuron, generalising the classical autoencoder in FIG.~\ref{fig:cl_autoencoder}. From each of these neurons, one output qubit each is led into the bottleneck neuron $U_1$, followed by a fan-out of its output. We add as an extra desideratum that the compressed bit, the output of $U_1$, is diagonal in the computational basis.  
The final neurons have the task of recreating $\ket{in_{12}}$ on the outputs labelled 6 and 8 respectively. 
The result is shown in FIG.~\ref{fig:quantum_autoencoder}. 
This means that a natural and simple cost function is  
\begin{equation}C=\sum_{j=0, k=0}^3 (\mathrm{Tr}(\rho_{6,8}\sigma_{j}\otimes \sigma_{k}) - \mathrm{Tr}(\rho_{in_{1,2}}\sigma_{j}\otimes \sigma_{k}))^{2}.
\end{equation}
Training is then conducted via global gradient descent of the cost
w.r.t. the $\alpha_{j_{1},...,j_{N}}$ parameters, as defined in
Eq.~\ref{eq:learningruleq}. During the training the network was fed states from the given input set, picked independently and identically for each step (i.i.d). 
 Standard speed-up techniques for learning
were used, e.g. a momentum term~\cite{Azoff94, Nielsen15}. In training with a variety of 2 possible orthogonal input states including superposition states, the cost function of the quantum autoencoder converged towards zero through global gradient descent in every case, starting with uniformly randomised weights, $\alpha_{j_{1},...,j_{N}} \in [-1,1]$.  For 2 non-orthogonal inputs and a 1-qubit bottleneck the cost-function will not converge to zero as is to be expected, but the training rather results in an approximately  compressing unitary.  
 FIG.~\ref{fig:quantum_autoencoder} shows the network learning to compress in the case of two possible inputs: $(\ket{00} +\ket{11})/\sqrt{2}$ and $(\ket{00} -
\ket{11})/\sqrt{2}$.
One can force the compressed output to be diagonal in a particular basis by adding an extra term to the cost-function (e.g.\ desiring the expectation value of Pauli X and Y to be zero in the case of a single qubit will push the network to give an output diagonal in the Z-basis).

\subsection*{Example: Neural network discovers teleportation protocol}
\begin{figure}[h]
\includegraphics[height=3.6cm]{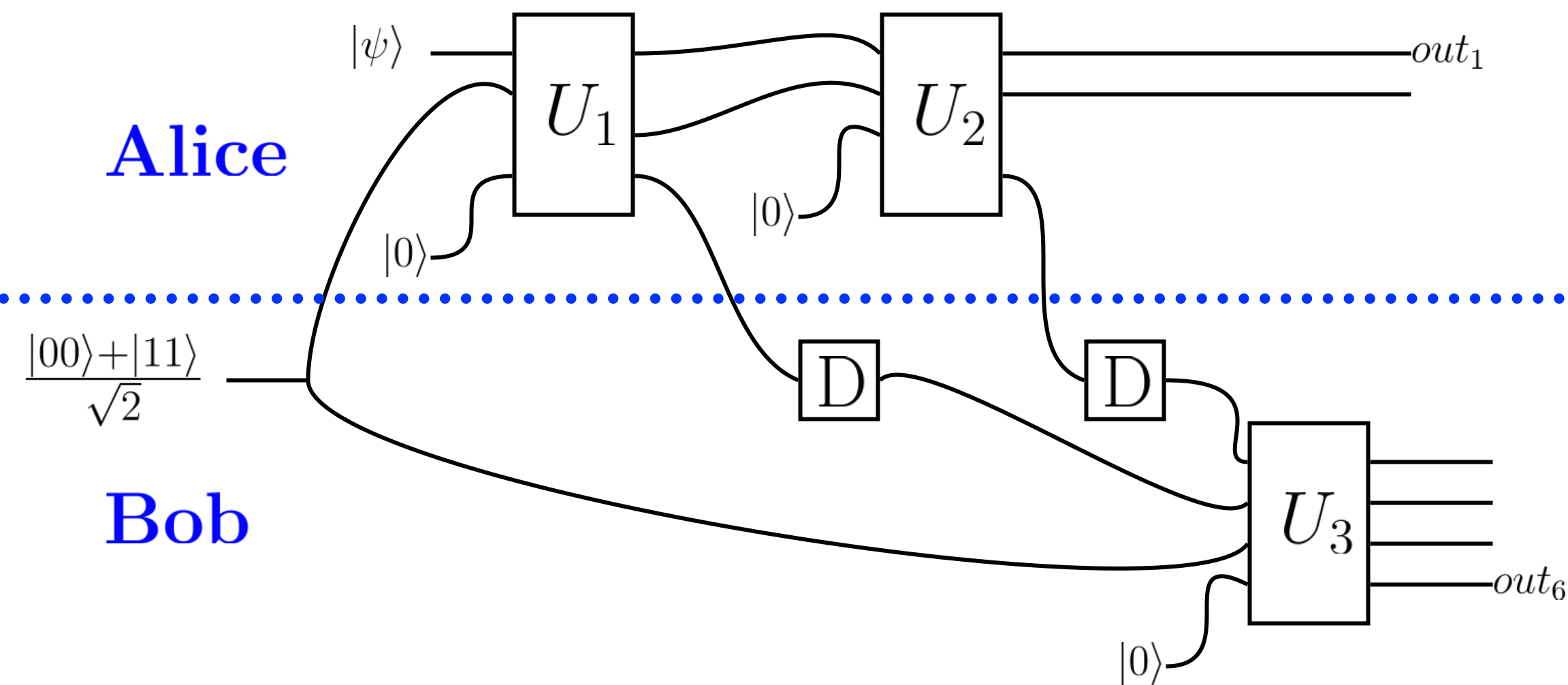}
\caption{\label{fig:teleport_circuit}A circuit diagram of a quantum
  neural network that can learn and carry out teleportation of the state
  $\ket{\psi}$ from Alice to Bob using quantum entanglement. The
  standard teleportation protocol allows only classical
  communication of 2 bits~\cite{NielsenChuang00}; this is enforced by only allowing two connections,  
  which are dephased in the Z-basis ($D$). $U_{1}, U_{2}$ and $U_{3}$ are
  unitaries,. The blue line is the boundary between Alice and Bob.}
\end{figure}

With quantum neural networks already shown to be able to perform
generalisations of classical tasks, we now consider the possibility of
quantum networks discovering solutions to existing and potentially
undiscovered quantum protocols. We propose a quantum neural network structure that can, on its own, work out
the standard protocol for quantum 
teleportation \cite{NielsenChuang00}. 

The design and training of this network is analogous to the
autoencoder and the quantum circuit diagram is shown in FIG.~\ref{fig:teleport_circuit}.
The cost function used was:
\begin{equation}
C=\sum_{j=0}^3(\mathrm{Tr}(\ket{\psi}\bra{\psi}\sigma_{j}) - \mathrm{Tr}(\rho_6 \sigma_{j}))^{2}.
\end{equation}
A fully
trained network can teleport the state $\ket{\psi}$ (from Alice) to the
output port of qubit 6 (to Bob). Once trained properly,
$\rho_{out_{1}}$ will no longer be
$\ket{\psi}\bra{\psi}$, as the teleportation has `messed up' Alice's state~\cite{Wilde13}. 

\begin{figure}[h]
\includegraphics[height=5cm]{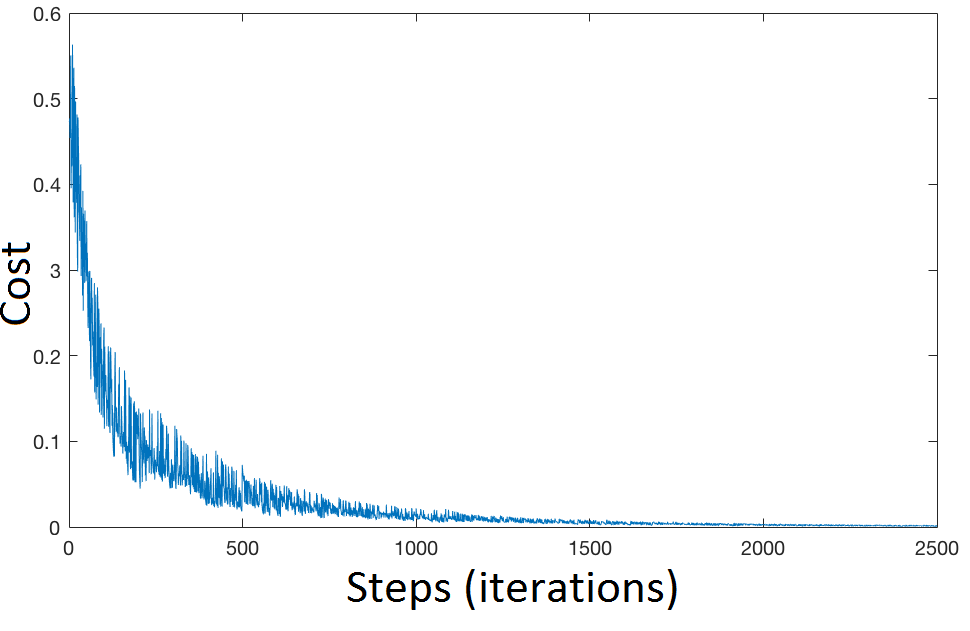}
\caption{\label{fig:teleportation_cost}A plot of the teleportation cost
  function w.r.t. the number of steps used in the training
  procedure. The cost function can be seen to converge to zero. The
  non-monotonic decrease is to be expected as we are varying the input
  states. The network now teleports any qubit state: picking 1000
  states at random from the Haar measure (uniform distribution over
  the Bloch sphere) gives a cost function distribution with mean
  $5.0371\times 10^{-4}$ and standard deviation $1.7802\times 10^{-4}$, which is effectively zero.}
\end{figure}

In order to train the teleportation for any arbitrary state $\ket{\psi}$
(and to avoid the network simply learning to \emph{copy} $\ket{\psi}$ from
Alice to Bob), the training inputs are randomly picked from the axis intersection states on the
surface of the Bloch sphere~\cite{NielsenChuang00}. FIG.~\ref{fig:teleportation_cost}
shows the convergence of the cost function during training, simulated
on a classical computer. As can be seen, the training was found to be
successful, i.e. the cost function converged towards zero. This held for all tests with randomly initialised weights.
\section*{Discussion}
\subsection*{Quantum vs. classical}
Can these neural networks show some form of quantum supremacy? The comparison of classical and quantum neural networks is well-defined within our set-up, as the classical networks correspond to a particular parameter regime for the quantum networks.  A key type of quantum supremacy is that the quantum network can take and process quantum inputs: it can for example process $\ket{+}$ and $\ket{-}$ differently. Thus, there are numerous quantum tasks it can do that the classical network cannot, including the two examples above. We anticipate that they will moreover, in some cases be able to process classical inputs faster, by turning them into superpositions---investigating this is a natural follow-on from this work. 

We also mention that we term our above design a {\em quantum} neural network with {\em classical learning parameters}, as the parameters in the unitaries are classical. It seems plausible that allowing these parameters to be in superpositions, whilst experimentally more challenging, could give further advantages.

Whilst adding the ancillary qubits ensures that the network is a strict generalisation of the classical network, it can of course be experimentally and numerically simpler to omit these. Then one would sacrifice performance in the classical mode of operation, and the network may not be as good as a classical network with the same number of neurons for all tasks.

\subsection*{Visualising the cost function landscape}
\begin{figure}[h]
\includegraphics[height=5.75cm]{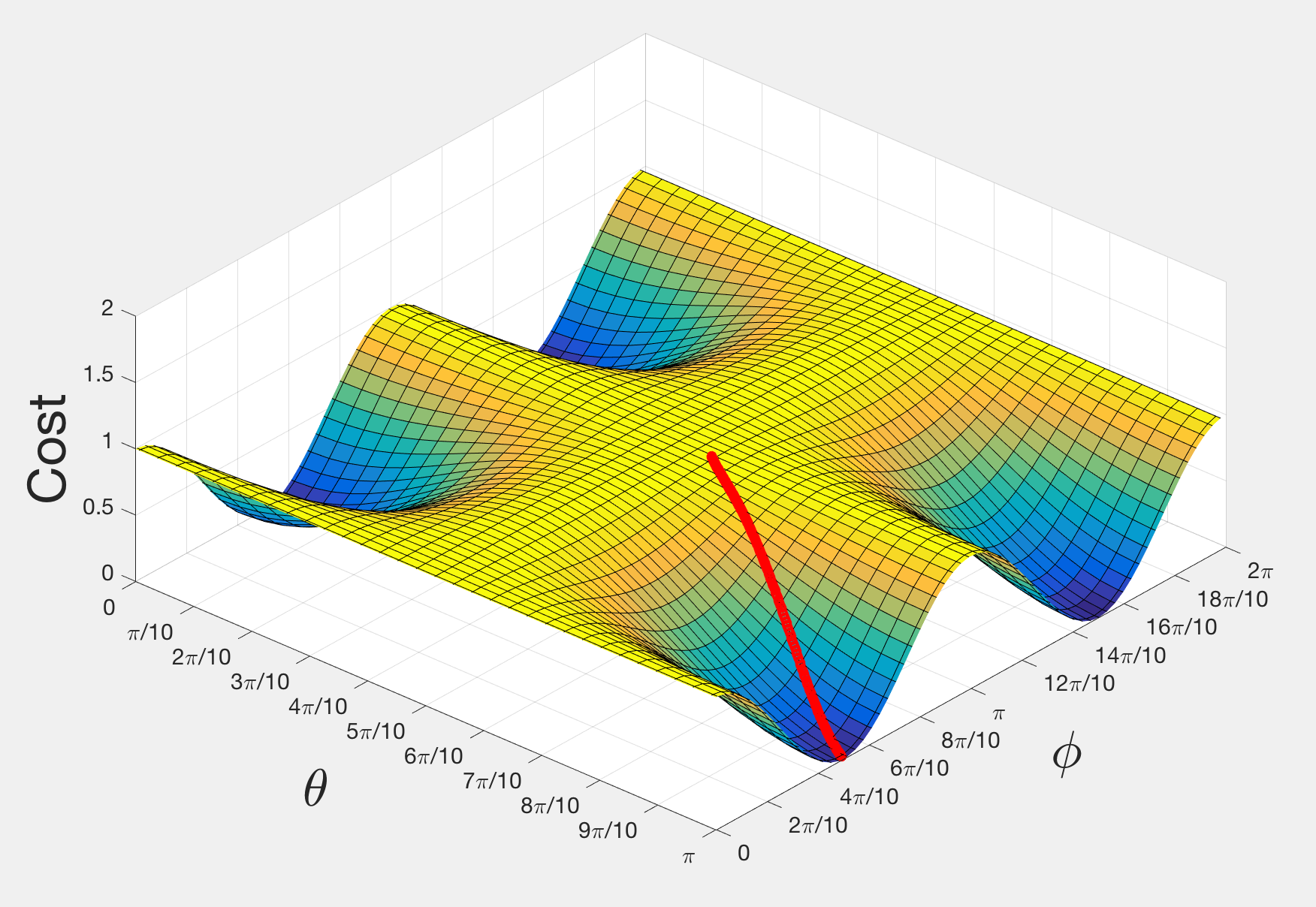}
\caption{\label{fig:grad_des_path}A 3-D plot of the cost function (vertical
  axis) of a 2-qubit unitary as a function of $\theta$ and $\phi$ (horizontal axes). The red line represents the
path taken when carrying out gradient descent from a particular
starting point.}
\end{figure}

To gain intuitive understanding, one can visualise the gradient descent in 3D by reducing the number of free parameters. We sampled the cost surface and gradient descent path of a one-input neuron ($4 \times4$ 
unitary matrix). With the second qubit expressed as the dummy-then-output qubit, the task for the neuron was $\ket{+
}\otimes\ket{0}\rightarrow
\ket{+}\otimes\ket{0}$ and $\ket{-}\otimes\ket{0} \rightarrow
\ket{-}\otimes\ket{1}$. We optimised, similarly to~Eq.~\ref{eq:u_nrestr}, over unitaries of the form 
\begin{equation}
U=\ket{\tau}\bra{\tau}\otimes \mathbb{1} + \ket{\tau^{\perp}}\bra{\tau^{\perp}}\otimes\sigma_{1},
\end{equation}
\noindent where $\ket{\tau} = \cos(\theta/2)\ket{0} +
e^{i\phi}\sin(\theta/2)\ket{1}$ 
 and $\ket{\tau^{\perp}} = \sin(\theta/2)\ket{0} -
e^{i\phi}\cos(\theta/2)\ket{1}$. We performed gradient descent along the variables $\theta$ and
$\phi$ as shown by the red path in FIG.~\ref{fig:grad_des_path}.

\subsection*{Scaling to bigger networks}
\begin{figure}[h]
\includegraphics[height=6cm]{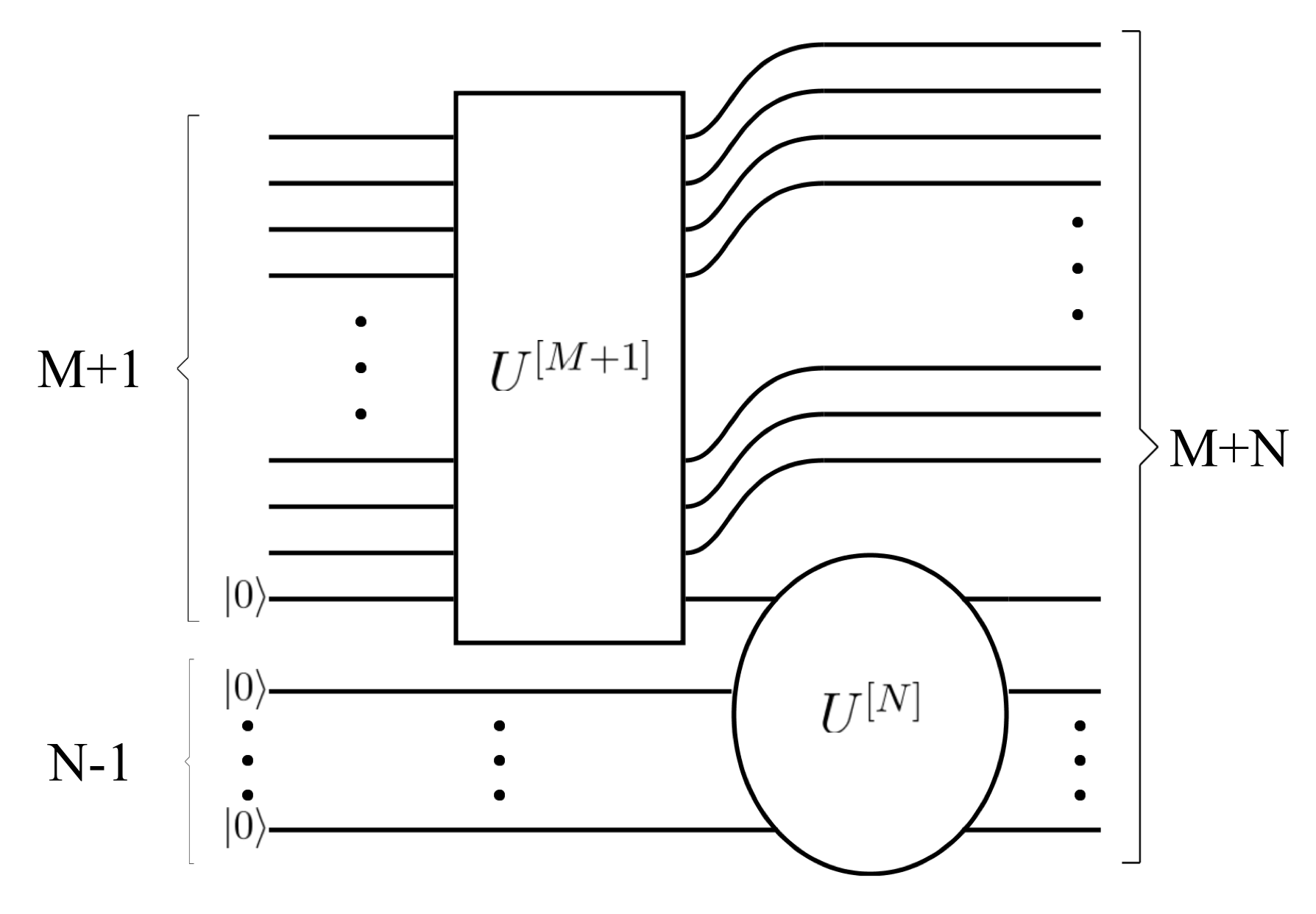}
\caption{\label{fig:general_neuron} Diagram of the quantum generalisation of a classical
  neuron with $M$ inputs and $N$ outputs. The superscripts inside the
  square brackets of the
  unitaries represent the number of qubits the respective unitaries
  act on. $U^{[M+1]}$ is the unitary that represents the quantum neuron
with an $N$-qubit input and $U^{[N]}$ is the fan-out gate that fans out the
output in the final port of $U^{[M+1]}$ in a particular basis.}
\end{figure}

The same scheme can be used to make quantum generalisations of networks whose generalised neurons have more inputs/outputs and connections.  FIG.~\ref{fig:general_neuron} illustrates an $M$-qubit input
quantum neuron with a subsequent $N$-qubit fan-out gate.

If one wishes the number of free parameters of a neuron to grow no more than polynomially in the number of inputs, one needs to restrict the unitary. It is natural to demand it to be 
a polynomial length circuit of some elementary universal gates, in particular if the input states are known to be generated by a polynomial length circuit of a given set of gates, it is natural to let the 
unitary be restricted to that set of gates.

The evaluation of the cost function can be kept to a sensible scaling if we restrict it to be a function of local observables on each qubit, in particular a function of the local Pauli expectation values, as was used in this paper, for which case a vector of $3n$ expectation values suffices for $n$ qubits.

\section*{Quantum photonics neuron module}
To investigate the physical viability of these quantum neural networks we consider quantum photonics. This is an  attractive platform for quantum information processing: it has room temperature operation, the possibility of robust miniaturisation through photonic integrated circuits; in general it harnesses the highly developed optical fibre-related technology for QIP purposes~\cite{silicon}. Moreover optical implementations have been viewed as optimal for neural networks, in the classical case, due to the low design cost of adding multiple connections (as light passes through light without interacting)~\cite{Rojas}. A final motivation for choosing this platform is that the tuning can be naturally implemented, as detailed below. 

We design a neuron as a module that can then be connected to other neurons. This makes it concrete how experimentally complex the network would be to build and operate, including how it could be trained.  

The design employs the Cerf-Adami-Kwiat (C-A-K) protocol~\cite{CAK}, where a single photon with polarisation and multiple possible spatial modes encodes the quantum state; the scheme falls into the category of hyper-entangling schemes, which entangle different degrees of freedom. One qubit is the polarisation; digital encodings of the spatial mode labels give rise to the others. With four spatial modes this implements 3 qubits, with basis $\ket{0/1}\ket{H/V}\ket{0/1}$, where $H/V$ are two different polarisation states, and the other bits label the four spatial modes. The first bit says whether it is in the top two or bottom two pairs of modes and the last bit whether it is the upper or lower one in one of those pairs. This scheme and related ones such as~\cite{Reck, Clements} are experimentally viable, theoretically clean and can implement any unitary on a single photon spread out over spatial modes. In such a single photon scenario they do not scale well however.
The number of spatial modes grows exponentially in the number of qubits. Thus for larger networks our design below would need to be modified to something less simple, e.g.\ accepting probabilistic gates in the spirit of the KLM scheme~\cite{klm}, or using measurement-based cluster state quantum computation approaches~\cite{silicon}.

\begin{figure}[h]
\includegraphics[height=2cm]{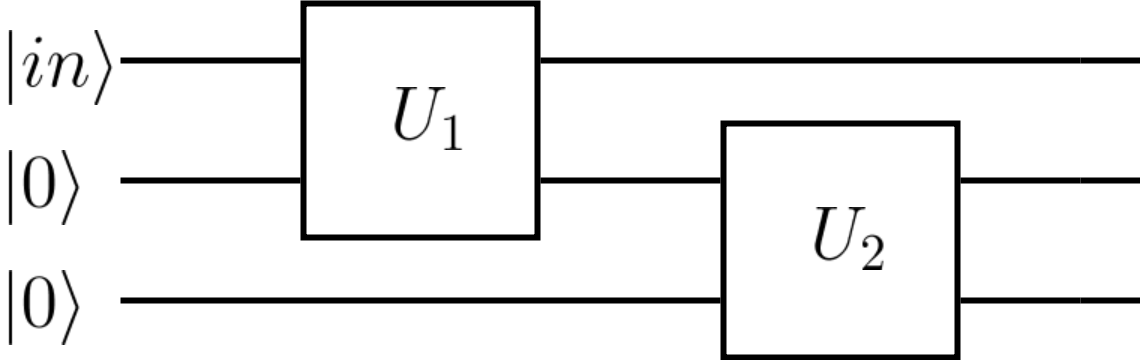}
\caption{\label{fig:onn} The first neuron takes one input and one dummy input 
and its designated output is fed into the next neuron.}
\end{figure}

\begin{figure}[h]
\includegraphics[width=\linewidth]{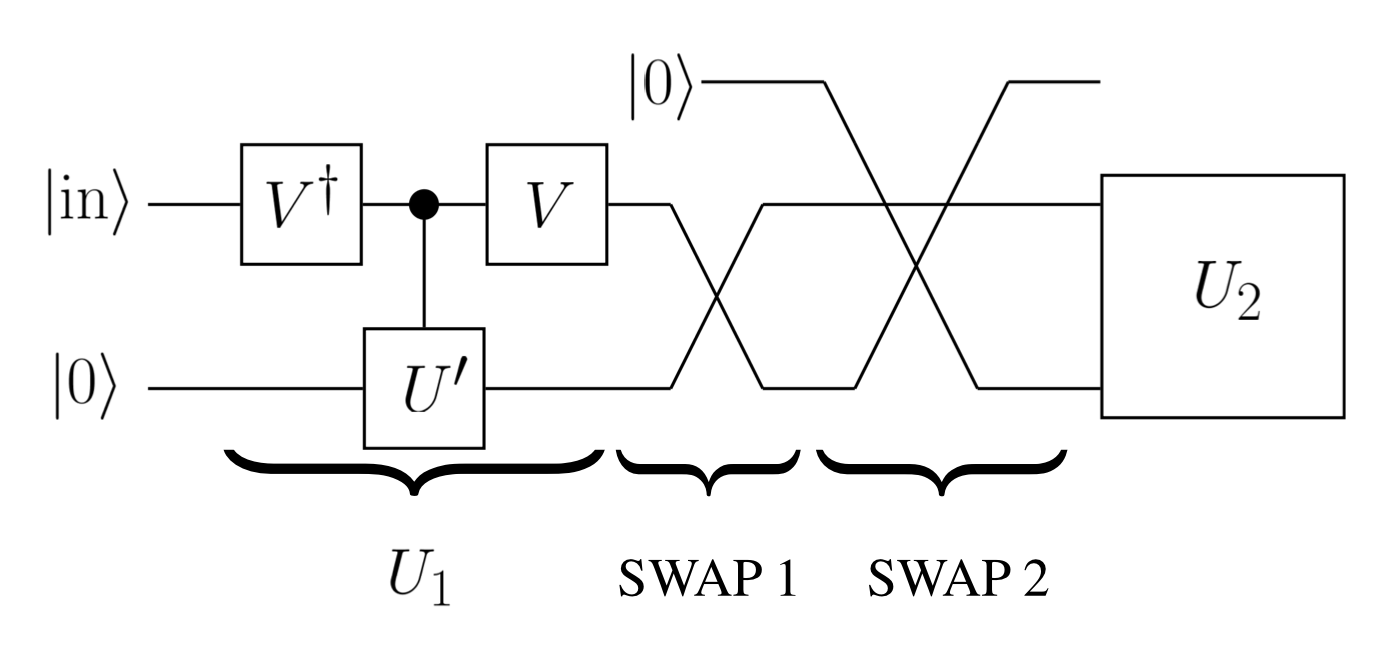}
\caption{\label{fig:ocircuit} A circuit diagram of our neural module.  Following C-A-K there are three qubits, with basis $\ket{0/1}\ket{H/V}\ket{0/1}$, where $H/V$ label different polarisation states, and the other bits label the four spatial modes. We define the input to the module to be carried by the middle (polarisation) qubit. The neuron $U_1$ has the form of Eq.~\ref{eq:u_nrestr}, modifying the output conditional on the input state. The swaps ensure that the next neuron module  $U_2$ also gets the input via the polarisation.}
\end{figure}

Before describing the module we make the simplifying restriction that there is one input qubit to the neuron and one dummy input. We will ensure that the designated output qubit can be fed into another neuron, as in FIG.~\ref{fig:onn} and FIG.~\ref{fig:ocircuit}.
\begin{figure*}[h]
\includegraphics[width=\linewidth]{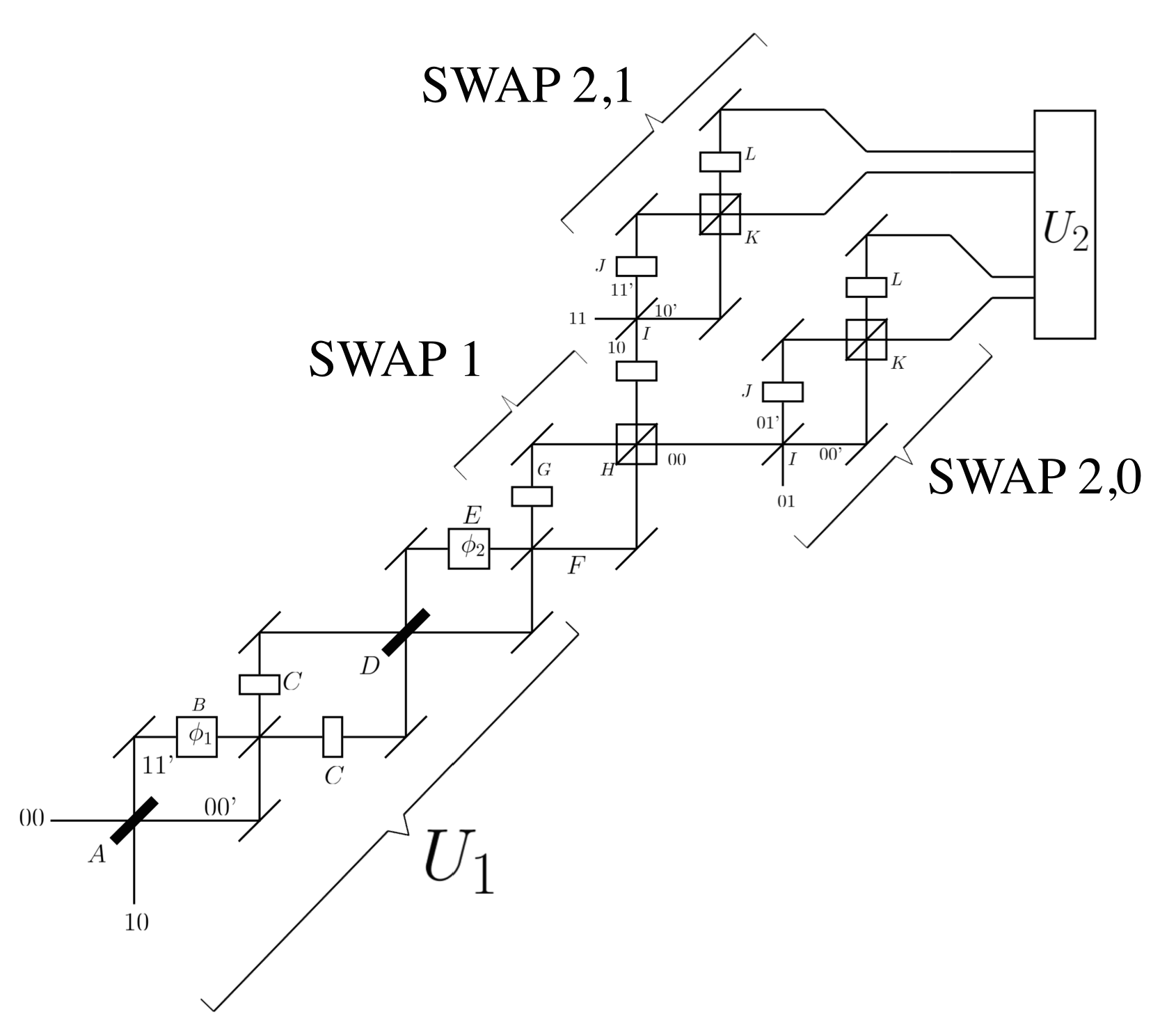}
\caption{\label{fig:optics_circuit} The optics circuit of the neuron module. There are four spatial modes labelled $\ket{00}, \ket{01}, \ket{10}$ and $\ket{11}$. Initially only $\ket{00}$ and $\ket{10}$ have non-zero amplitudes and the second spatial qubit is not manipulated. The polarisation of the single photon is also manipulated. The two beamsplitters in bold at points A and D are variable (and can be replaced by Mach-Zehnder interferometers with variable phase). B and E are variable phase shifters and  C shows a variable polarisation shifter. G and F are the two spacial modes available before a splitting occurs at H via a polarising beamsplitter, where the (fixed) polarisation rotator implements SWAP1. The beamsplitters with extra inputs at I allow for an additional spacial qubit to be manipulated, with J, K and L representing the components required for a SWAP gate. Before entering the second unitary, the ‘second level splitting’ modes are brought close.}
\end{figure*}
We propose to update the neural network by adjusting both variable polarisation rotators, and spatial phase shifters in a set of Mach-Zehnder interferometers as shown in FIG.~\ref{fig:optics_circuit}. In this we are able to change the outputs from each layer of the network. The spatial shift could be induced
by varying the strain or temperature on the waveguides at given locations, to change
their refractive indices and hence the relative phase; this may have
additional difficulties in that silicon waveguides are birefringent
~\cite{strain}. Alternatively we can tune both polarisation and spatial
qubits via the electro-optic effect.

This circuit can be made more robust and minitaturised using silicon or silica optical waveguides~\cite{silicon}. They have been extensively used to control spatial modes and recently also polarisation~\cite{polarisationstates}. Several labs can implement the phase shifting via heaters or the electro-optic effect. Conventionally phase shifters built upon the electro-optic effect are known to work in the megahertz region and have extremely low loss~\cite{silicon}. For many applications this would be considered slow, but our tuning only requires (in the region of) a few thousand steps of tuning, meaning learning tasks for neural networks this small could be completed in milliseconds. While it appears that this effect will be the limiting factor in terms of speed, photodetectors are able to reach reset times in the tens of nanoseconds, while the production of single photons through parametric down conversion have megahertz repetion rates~\cite{polarisationshift}. 
\section*{Summary and Outlook}
We have given a protocol for generalising classical feedforward step-function neural networks to networks that take and process quantum inputs. We have shown that these networks can perform the natural quantum generalisation 
of the classical network in the case of an autoencoder, being able for example to compress entangled inputs. We have shown that 
they can be used to work out a quantum information processing protocol: teleportation, without being told how to do it, only the task.
Based on these results we think that these networks will be highly versatile tools for 
quantum information scientists, similar to the classical networks' role in classical information processing. 
\section*{Acknowledgments}
We acknowledge discussions with Stefanie Baerz, Abbas Edalat, William Clements, Alex Jones, Mio Murao, Maria Schuld, Vlatko Vedral, Alejandro Valido and discussions and detailed comments from Doug Plato, Mihai Vidrighin, Peter Wittek. We are grateful for funding from the the EU Collaborative Project TherMiQ (Grant Agreement 618074), the London Institute for Mathematical Sciences, a Leverhulme Trust Research Grant
(No. RPG-2014-055), a programme grant from the UK EPSRC (EP/K034480/1).

\newpage

\bibliography{bibliography_bibtex}

\begin{thebibliography}{44}%
\makeatletter
\providecommand \@ifxundefined [1]{%
 \@ifx{#1\undefined}
}%
\providecommand \@ifnum [1]{%
 \ifnum #1\expandafter \@firstoftwo
 \else \expandafter \@secondoftwo
 \fi
}%
\providecommand \@ifx [1]{%
 \ifx #1\expandafter \@firstoftwo
 \else \expandafter \@secondoftwo
 \fi
}%
\providecommand \natexlab [1]{#1}%
\providecommand \enquote  [1]{``#1''}%
\providecommand \bibnamefont  [1]{#1}%
\providecommand \bibfnamefont [1]{#1}%
\providecommand \citenamefont [1]{#1}%
\providecommand \href@noop [0]{\@secondoftwo}%
\providecommand \href [0]{\begingroup \@sanitize@url \@href}%
\providecommand \@href[1]{\@@startlink{#1}\@@href}%
\providecommand \@@href[1]{\endgroup#1\@@endlink}%
\providecommand \@sanitize@url [0]{\catcode `\\12\catcode `\$12\catcode
  `\&12\catcode `\#12\catcode `\^12\catcode `\_12\catcode `\%12\relax}%
\providecommand \@@startlink[1]{}%
\providecommand \@@endlink[0]{}%
\providecommand \url  [0]{\begingroup\@sanitize@url \@url }%
\providecommand \@url [1]{\endgroup\@href {#1}{\urlprefix }}%
\providecommand \urlprefix  [0]{URL }%
\providecommand \Eprint [0]{\href }%
\providecommand \doibase [0]{http://dx.doi.org/}%
\providecommand \selectlanguage [0]{\@gobble}%
\providecommand \bibinfo  [0]{\@secondoftwo}%
\providecommand \bibfield  [0]{\@secondoftwo}%
\providecommand \translation [1]{[#1]}%
\providecommand \BibitemOpen [0]{}%
\providecommand \bibitemStop [0]{}%
\providecommand \bibitemNoStop [0]{.\EOS\space}%
\providecommand \EOS [0]{\spacefactor3000\relax}%
\providecommand \BibitemShut  [1]{\csname bibitem#1\endcsname}%
\let\auto@bib@innerbib\@empty
\bibitem [{\citenamefont {Nielsen}(1991)}]{Nielsen15}%
  \BibitemOpen
  \bibfield  {author} {\bibinfo {author} {\bibfnamefont {M.~A.}\ \bibnamefont
  {Nielsen}},\ }\href@noop {} {\emph {\bibinfo {title} {Neural Networks and
  Deep Learning}}}\ (\bibinfo  {publisher} {Determination Press},\ \bibinfo
  {address} {online book},\ \bibinfo {year} {1991})\BibitemShut {NoStop}%
\bibitem [{\citenamefont {Azoff}(1994)}]{Azoff94}%
  \BibitemOpen
  \bibfield  {author} {\bibinfo {author} {\bibfnamefont {E.~M.}\ \bibnamefont
  {Azoff}},\ }\href@noop {} {\emph {\bibinfo {title} {Neural Network Time
  Series Forecasting of Financial Markets}}}\ (\bibinfo  {publisher} {John
  Wiley and Sons},\ \bibinfo {address} {Chichester},\ \bibinfo {year}
  {1994})\BibitemShut {NoStop}%
\bibitem [{\citenamefont {LeCunn}\ \emph {et~al.}(2015)\citenamefont {LeCunn},
  \citenamefont {Bengio},\ and\ \citenamefont {Hinton}}]{LecunBengioHinton15}%
  \BibitemOpen
  \bibfield  {author} {\bibinfo {author} {\bibfnamefont {Y.}~\bibnamefont
  {LeCunn}}, \bibinfo {author} {\bibfnamefont {Y.}~\bibnamefont {Bengio}}, \
  and\ \bibinfo {author} {\bibfnamefont {G.}~\bibnamefont {Hinton}},\
  }\bibfield  {title} {\enquote {\bibinfo {title} {Deep learning},}\
  }\href@noop {} {\bibfield  {journal} {\bibinfo  {journal} {Nature}\ }\textbf
  {\bibinfo {volume} {521}},\ \bibinfo {pages} {436--444} (\bibinfo {year}
  {2015})}\BibitemShut {NoStop}%
\bibitem [{\citenamefont {Lloyd}\ \emph {et~al.}(2013)\citenamefont {Lloyd},
  \citenamefont {Mohseni},\ and\ \citenamefont {Rebentrost}}]{LloydMR13}%
  \BibitemOpen
  \bibfield  {author} {\bibinfo {author} {\bibfnamefont {S.}~\bibnamefont
  {Lloyd}}, \bibinfo {author} {\bibfnamefont {M.}~\bibnamefont {Mohseni}}, \
  and\ \bibinfo {author} {\bibfnamefont {P.}~\bibnamefont {Rebentrost}},\
  }\bibfield  {title} {\enquote {\bibinfo {title} {Quantum algorithms for
  supervised and unsupervised machine learning},}\ }\href@noop {} {\  (\bibinfo
  {year} {2013})},\ \Eprint {http://arxiv.org/abs/1307.0411} {arXiv:1307.0411
  [quant-ph]} \BibitemShut {NoStop}%
\bibitem [{\citenamefont {Lloyd}\ \emph {et~al.}(2014)\citenamefont {Lloyd},
  \citenamefont {Mohseni},\ and\ \citenamefont {Rebentrost}}]{LloydMR13ii}%
  \BibitemOpen
  \bibfield  {author} {\bibinfo {author} {\bibfnamefont {S.}~\bibnamefont
  {Lloyd}}, \bibinfo {author} {\bibfnamefont {M.}~\bibnamefont {Mohseni}}, \
  and\ \bibinfo {author} {\bibfnamefont {P.}~\bibnamefont {Rebentrost}},\
  }\bibfield  {title} {\enquote {\bibinfo {title} {Quantum principal component
  analysis},}\ }\href@noop {} {\bibfield  {journal} {\bibinfo  {journal}
  {Nature Physics}\ }\textbf {\bibinfo {volume} {10}},\ \bibinfo {pages}
  {631–--633} (\bibinfo {year} {2014})}\BibitemShut {NoStop}%
\bibitem [{\citenamefont {Montanaro}(2015)}]{Montanaro15}%
  \BibitemOpen
  \bibfield  {author} {\bibinfo {author} {\bibfnamefont {A.}~\bibnamefont
  {Montanaro}},\ }\bibfield  {title} {\enquote {\bibinfo {title} {Quantum
  pattern matching fast on average},}\ }\href {\doibase
  10.1007/s00453-015-0060-4} {\bibfield  {journal} {\bibinfo  {journal}
  {Algorithmica}\ ,\ \bibinfo {pages} {1--24}} (\bibinfo {year}
  {2015})}\BibitemShut {NoStop}%
\bibitem [{\citenamefont {Aaronson}(2015)}]{Aaronson15}%
  \BibitemOpen
  \bibfield  {author} {\bibinfo {author} {\bibfnamefont {S.}~\bibnamefont
  {Aaronson}},\ }\bibfield  {title} {\enquote {\bibinfo {title} {Read the fine
  print},}\ }\href {http://dx.doi.org/10.1038/nphys3272} {\bibfield  {journal}
  {\bibinfo  {journal} {Nature Physics}\ }\textbf {\bibinfo {volume} {11}},\
  \bibinfo {pages} {291--293} (\bibinfo {year} {2015})}\BibitemShut {NoStop}%
\bibitem [{\citenamefont {Garnerone}\ \emph {et~al.}(2012)\citenamefont
  {Garnerone}, \citenamefont {Zanardi},\ and\ \citenamefont
  {Lidar}}]{GarneroneZL12}%
  \BibitemOpen
  \bibfield  {author} {\bibinfo {author} {\bibfnamefont {S.}~\bibnamefont
  {Garnerone}}, \bibinfo {author} {\bibfnamefont {P.}~\bibnamefont {Zanardi}},
  \ and\ \bibinfo {author} {\bibfnamefont {D.~A.}\ \bibnamefont {Lidar}},\
  }\bibfield  {title} {\enquote {\bibinfo {title} {Adiabatic quantum algorithm
  for search engine ranking},}\ }\href {\doibase
  10.1103/PhysRevLett.108.230506} {\bibfield  {journal} {\bibinfo  {journal}
  {Phys. Rev. Lett.}\ }\textbf {\bibinfo {volume} {108}},\ \bibinfo {pages}
  {230506} (\bibinfo {year} {2012})}\BibitemShut {NoStop}%
\bibitem [{\citenamefont {Harrow}\ \emph {et~al.}(2009)\citenamefont {Harrow},
  \citenamefont {Hassidim},\ and\ \citenamefont {Lloyd}}]{HarrowHL09}%
  \BibitemOpen
  \bibfield  {author} {\bibinfo {author} {\bibfnamefont {A.~W.}\ \bibnamefont
  {Harrow}}, \bibinfo {author} {\bibfnamefont {A.}~\bibnamefont {Hassidim}}, \
  and\ \bibinfo {author} {\bibfnamefont {S.}~\bibnamefont {Lloyd}},\ }\bibfield
   {title} {\enquote {\bibinfo {title} {Quantum algorithm for linear systems of
  equations},}\ }\href {\doibase 10.1103/PhysRevLett.103.150502} {\bibfield
  {journal} {\bibinfo  {journal} {Phys. Rev. Lett.}\ }\textbf {\bibinfo
  {volume} {103}},\ \bibinfo {pages} {150502} (\bibinfo {year}
  {2009})}\BibitemShut {NoStop}%
\bibitem [{\citenamefont {Lloyd}\ \emph {et~al.}(2016)\citenamefont {Lloyd},
  \citenamefont {Garnerone},\ and\ \citenamefont {Zanardi}}]{LloydGZ16}%
  \BibitemOpen
  \bibfield  {author} {\bibinfo {author} {\bibfnamefont {S.}~\bibnamefont
  {Lloyd}}, \bibinfo {author} {\bibfnamefont {S.}~\bibnamefont {Garnerone}}, \
  and\ \bibinfo {author} {\bibfnamefont {P.}~\bibnamefont {Zanardi}},\
  }\bibfield  {title} {\enquote {\bibinfo {title} {Quantum algorithms for
  topological and geometric analysis of big data},}\ }\href {\doibase
  10.1038/ncomms10138} {\bibfield  {journal} {\bibinfo  {journal} {Nature
  Communications}\ }\textbf {\bibinfo {volume} {7}},\ \bibinfo {pages} {10138}
  (\bibinfo {year} {2016})}\BibitemShut {NoStop}%
\bibitem [{\citenamefont {Rebentrost}\ \emph {et~al.}(2014)\citenamefont
  {Rebentrost}, \citenamefont {Mohseni},\ and\ \citenamefont
  {Lloyd}}]{RebenstrostML13}%
  \BibitemOpen
  \bibfield  {author} {\bibinfo {author} {\bibfnamefont {P.}~\bibnamefont
  {Rebentrost}}, \bibinfo {author} {\bibfnamefont {M.}~\bibnamefont {Mohseni}},
  \ and\ \bibinfo {author} {\bibfnamefont {S.}~\bibnamefont {Lloyd}},\
  }\bibfield  {title} {\enquote {\bibinfo {title} {Quantum support vector
  machine for big data classification},}\ }\href {\doibase
  10.1103/PhysRevLett.113.130503} {\bibfield  {journal} {\bibinfo  {journal}
  {Phys. Rev. Lett.}\ }\textbf {\bibinfo {volume} {113}},\ \bibinfo {pages}
  {130503} (\bibinfo {year} {2014})}\BibitemShut {NoStop}%
\bibitem [{\citenamefont {Wiebe}\ \emph {et~al.}(2012)\citenamefont {Wiebe},
  \citenamefont {Braun},\ and\ \citenamefont {Lloyd}}]{WiebeBL12}%
  \BibitemOpen
  \bibfield  {author} {\bibinfo {author} {\bibfnamefont {N.}~\bibnamefont
  {Wiebe}}, \bibinfo {author} {\bibfnamefont {D.}~\bibnamefont {Braun}}, \ and\
  \bibinfo {author} {\bibfnamefont {S.}~\bibnamefont {Lloyd}},\ }\bibfield
  {title} {\enquote {\bibinfo {title} {Quantum algorithm for data fitting},}\
  }\href {\doibase 10.1103/PhysRevLett.109.050505} {\bibfield  {journal}
  {\bibinfo  {journal} {Phys. Rev. Lett.}\ }\textbf {\bibinfo {volume} {109}},\
  \bibinfo {pages} {050505} (\bibinfo {year} {2012})}\BibitemShut {NoStop}%
\bibitem [{\citenamefont {Adcock}\ \emph {et~al.}(2015)\citenamefont {Adcock},
  \citenamefont {Allen}, \citenamefont {Day}, \citenamefont {Frick},
  \citenamefont {Hinchliff}, \citenamefont {Johnson}, \citenamefont
  {Morley-Short}, \citenamefont {Pallister}, \citenamefont {Price},\ and\
  \citenamefont {Stanisic}}]{Adcock15}%
  \BibitemOpen
  \bibfield  {author} {\bibinfo {author} {\bibfnamefont {J.}~\bibnamefont
  {Adcock}}, \bibinfo {author} {\bibfnamefont {E.}~\bibnamefont {Allen}},
  \bibinfo {author} {\bibfnamefont {M.}~\bibnamefont {Day}}, \bibinfo {author}
  {\bibfnamefont {S.}~\bibnamefont {Frick}}, \bibinfo {author} {\bibfnamefont
  {J.}~\bibnamefont {Hinchliff}}, \bibinfo {author} {\bibfnamefont
  {M.}~\bibnamefont {Johnson}}, \bibinfo {author} {\bibfnamefont
  {S.}~\bibnamefont {Morley-Short}}, \bibinfo {author} {\bibfnamefont
  {S.}~\bibnamefont {Pallister}}, \bibinfo {author} {\bibfnamefont
  {A.}~\bibnamefont {Price}}, \ and\ \bibinfo {author} {\bibfnamefont
  {S.}~\bibnamefont {Stanisic}},\ }\bibfield  {title} {\enquote {\bibinfo
  {title} {Advances in quantum machine learning},}\ }\href@noop {} {\
  (\bibinfo {year} {2015})},\ \Eprint {http://arxiv.org/abs/1512.02900}
  {arXiv:1512.02900 [quant-ph]} \BibitemShut {NoStop}%
\bibitem [{\citenamefont {Heim}\ \emph {et~al.}(2015)\citenamefont {Heim},
  \citenamefont {R{\o}nnow}, \citenamefont {Isakov},\ and\ \citenamefont
  {Troyer}}]{HeimRIT15}%
  \BibitemOpen
  \bibfield  {author} {\bibinfo {author} {\bibfnamefont {B.}~\bibnamefont
  {Heim}}, \bibinfo {author} {\bibfnamefont {T.~F.}\ \bibnamefont {R{\o}nnow}},
  \bibinfo {author} {\bibfnamefont {S.~V.}\ \bibnamefont {Isakov}}, \ and\
  \bibinfo {author} {\bibfnamefont {M.}~\bibnamefont {Troyer}},\ }\bibfield
  {title} {\enquote {\bibinfo {title} {Quantum versus classical annealing of
  {I}sing spin glasses},}\ }\href {\doibase 10.1126/science.aaa4170} {\bibfield
   {journal} {\bibinfo  {journal} {Science}\ }\textbf {\bibinfo {volume}
  {348}},\ \bibinfo {pages} {215--217} (\bibinfo {year} {2015})}\BibitemShut
  {NoStop}%
\bibitem [{\citenamefont {Gross}\ \emph {et~al.}(2010)\citenamefont {Gross},
  \citenamefont {Liu}, \citenamefont {Flammia}, \citenamefont {Becker},\ and\
  \citenamefont {Eisert}}]{GrossYFBE10}%
  \BibitemOpen
  \bibfield  {author} {\bibinfo {author} {\bibfnamefont {D.}~\bibnamefont
  {Gross}}, \bibinfo {author} {\bibfnamefont {Y.K.}\ \bibnamefont {Liu}},
  \bibinfo {author} {\bibfnamefont {S.~T.}\ \bibnamefont {Flammia}}, \bibinfo
  {author} {\bibfnamefont {S.}~\bibnamefont {Becker}}, \ and\ \bibinfo {author}
  {\bibfnamefont {J.}~\bibnamefont {Eisert}},\ }\bibfield  {title} {\enquote
  {\bibinfo {title} {Quantum state tomography via compressed sensing},}\ }\href
  {\doibase 10.1103/PhysRevLett.105.150401} {\bibfield  {journal} {\bibinfo
  {journal} {Phys. Rev. Lett.}\ }\textbf {\bibinfo {volume} {105}},\ \bibinfo
  {pages} {150401} (\bibinfo {year} {2010})}\BibitemShut {NoStop}%
\bibitem [{\citenamefont {Dunjko}\ \emph {et~al.}(2016)\citenamefont {Dunjko},
  \citenamefont {Taylor},\ and\ \citenamefont {Briegel}}]{Dunjko16}%
  \BibitemOpen
  \bibfield  {author} {\bibinfo {author} {\bibfnamefont {V.}~\bibnamefont
  {Dunjko}}, \bibinfo {author} {\bibfnamefont {J.~M.}\ \bibnamefont {Taylor}},
  \ and\ \bibinfo {author} {\bibfnamefont {H.~J.}\ \bibnamefont {Briegel}},\
  }\bibfield  {title} {\enquote {\bibinfo {title} {Quantum-enhanced machine
  learning},}\ }\href {\doibase 10.1103/PhysRevLett.117.130501} {\bibfield
  {journal} {\bibinfo  {journal} {Phys. Rev. Lett.}\ }\textbf {\bibinfo
  {volume} {117}},\ \bibinfo {pages} {130501} (\bibinfo {year}
  {2016})}\BibitemShut {NoStop}%
\bibitem [{\citenamefont {Wittek}(2014)}]{Wittek14}%
  \BibitemOpen
  \bibinfo {editor} {\bibfnamefont {P.}~\bibnamefont {Wittek}},\ ed.,\ \href
  {\doibase http://dx.doi.org/10.1016/B978-0-12-800953-6.00015-3} {\emph
  {\bibinfo {title} {Quantum Machine Learning}}}\ (\bibinfo  {publisher}
  {Academic Press},\ \bibinfo {address} {Boston},\ \bibinfo {year} {2014})\
  pp.\ \bibinfo {pages} {i -- ii}\BibitemShut {NoStop}%
\bibitem [{\citenamefont {Nielsen}\ and\ \citenamefont
  {Chuang}(2000)}]{NielsenChuang00}%
  \BibitemOpen
  \bibfield  {author} {\bibinfo {author} {\bibfnamefont {M.~A.}\ \bibnamefont
  {Nielsen}}\ and\ \bibinfo {author} {\bibfnamefont {I.~L.}\ \bibnamefont
  {Chuang}},\ }\href@noop {} {\emph {\bibinfo {title} {Quantum Computation and
  Quantum Information}}}\ (\bibinfo  {publisher} {Cambridge University Press},\
  \bibinfo {address} {Cambridge},\ \bibinfo {year} {2000})\BibitemShut
  {NoStop}%
\bibitem [{\citenamefont {Garner}\ \emph {et~al.}(2013)\citenamefont {Garner},
  \citenamefont {Dahlsten}, \citenamefont {Nakata}, \citenamefont {Murao},\
  and\ \citenamefont {Vedral}}]{GarnerDNMV15}%
  \BibitemOpen
  \bibfield  {author} {\bibinfo {author} {\bibfnamefont {A.~J.~P.}\
  \bibnamefont {Garner}}, \bibinfo {author} {\bibfnamefont {O.~C.~O.}\
  \bibnamefont {Dahlsten}}, \bibinfo {author} {\bibfnamefont {Y.}~\bibnamefont
  {Nakata}}, \bibinfo {author} {\bibfnamefont {M.}~\bibnamefont {Murao}}, \
  and\ \bibinfo {author} {\bibfnamefont {V.}~\bibnamefont {Vedral}},\
  }\bibfield  {title} {\enquote {\bibinfo {title} {A framework for phase and
  interference in generalized probabilistic theories},}\ }\href
  {http://stacks.iop.org/1367-2630/15/i=9/a=093044} {\bibfield  {journal}
  {\bibinfo  {journal} {New Journal of Physics}\ }\textbf {\bibinfo {volume}
  {15}},\ \bibinfo {pages} {093044} (\bibinfo {year} {2013})}\BibitemShut
  {NoStop}%
\bibitem [{\citenamefont {Lechner}\ \emph {et~al.}(2015)\citenamefont
  {Lechner}, \citenamefont {Hauke},\ and\ \citenamefont
  {Zoller}}]{Lechnere1500838}%
  \BibitemOpen
  \bibfield  {author} {\bibinfo {author} {\bibfnamefont {W.}~\bibnamefont
  {Lechner}}, \bibinfo {author} {\bibfnamefont {P.}~\bibnamefont {Hauke}}, \
  and\ \bibinfo {author} {\bibfnamefont {P.}~\bibnamefont {Zoller}},\
  }\bibfield  {title} {\enquote {\bibinfo {title} {A quantum annealing
  architecture with all-to-all connectivity from local interactions},}\
  }\href@noop {} {\bibfield  {journal} {\bibinfo  {journal} {Science Advances}\
  }\textbf {\bibinfo {volume} {1}} (\bibinfo {year} {2015})}\BibitemShut
  {NoStop}%
\bibitem [{\citenamefont {Wiebe}\ \emph {et~al.}(2014)\citenamefont {Wiebe},
  \citenamefont {Kapoor},\ and\ \citenamefont {Svore}}]{quantum-deep-learning}%
  \BibitemOpen
  \bibfield  {author} {\bibinfo {author} {\bibfnamefont {N.}~\bibnamefont
  {Wiebe}}, \bibinfo {author} {\bibfnamefont {A.}~\bibnamefont {Kapoor}}, \
  and\ \bibinfo {author} {\bibfnamefont {K.~M.}\ \bibnamefont {Svore}},\ }\href
  {https://www.microsoft.com/en-us/research/publication/quantum-deep-learning/}
  {\enquote {\bibinfo {title} {Quantum deep learning},}\ } (\bibinfo {year}
  {2014}),\ \bibinfo {note} {unpublished}\BibitemShut {NoStop}%
\bibitem [{\citenamefont {Schuld}\ \emph {et~al.}(2014)\citenamefont {Schuld},
  \citenamefont {Sinayskiy},\ and\ \citenamefont {Petruccione}}]{Schuld14}%
  \BibitemOpen
  \bibfield  {author} {\bibinfo {author} {\bibfnamefont {M.}~\bibnamefont
  {Schuld}}, \bibinfo {author} {\bibfnamefont {I.}~\bibnamefont {Sinayskiy}}, \
  and\ \bibinfo {author} {\bibfnamefont {F.}~\bibnamefont {Petruccione}},\
  }\bibfield  {title} {\enquote {\bibinfo {title} {The quest for a quantum
  neural network},}\ }\href@noop {} {\bibfield  {journal} {\bibinfo  {journal}
  {Quantum Information Processing}\ }\textbf {\bibinfo {volume} {13}},\
  \bibinfo {pages} {2567–2586} (\bibinfo {year} {2014})}\BibitemShut
  {NoStop}%
\bibitem [{\citenamefont {Bisio}\ \emph {et~al.}(2010)\citenamefont {Bisio},
  \citenamefont {Chiribella}, \citenamefont {D'Ariano}, \citenamefont
  {Facchini},\ and\ \citenamefont {Perinotti}}]{Bisio10}%
  \BibitemOpen
  \bibfield  {author} {\bibinfo {author} {\bibfnamefont {A.}~\bibnamefont
  {Bisio}}, \bibinfo {author} {\bibfnamefont {G.}~\bibnamefont {Chiribella}},
  \bibinfo {author} {\bibfnamefont {G.~M.}\ \bibnamefont {D'Ariano}}, \bibinfo
  {author} {\bibfnamefont {S.}~\bibnamefont {Facchini}}, \ and\ \bibinfo
  {author} {\bibfnamefont {P.}~\bibnamefont {Perinotti}},\ }\bibfield  {title}
  {\enquote {\bibinfo {title} {Optimal quantum learning of a unitary
  transformation},}\ }\href {\doibase 10.1103/PhysRevA.81.032324} {\bibfield
  {journal} {\bibinfo  {journal} {Phys. Rev. A}\ }\textbf {\bibinfo {volume}
  {81}},\ \bibinfo {pages} {032324} (\bibinfo {year} {2010})}\BibitemShut
  {NoStop}%
\bibitem [{\citenamefont {Sasaki}\ and\ \citenamefont
  {Carlini}(2002)}]{Sasaki02}%
  \BibitemOpen
  \bibfield  {author} {\bibinfo {author} {\bibfnamefont {M.}~\bibnamefont
  {Sasaki}}\ and\ \bibinfo {author} {\bibfnamefont {A.}~\bibnamefont
  {Carlini}},\ }\bibfield  {title} {\enquote {\bibinfo {title} {Quantum
  learning and universal quantum matching machine},}\ }\href {\doibase
  10.1103/PhysRevA.66.022303} {\bibfield  {journal} {\bibinfo  {journal} {Phys.
  Rev. A}\ }\textbf {\bibinfo {volume} {66}},\ \bibinfo {pages} {022303}
  (\bibinfo {year} {2002})}\BibitemShut {NoStop}%
\bibitem [{\citenamefont {Sent{\'i}s}\ \emph {et~al.}(2015)\citenamefont
  {Sent{\'i}s}, \citenamefont {Gu{\c{T}}{\u{a}}},\ and\ \citenamefont
  {Adesso}}]{Sentis15}%
  \BibitemOpen
  \bibfield  {author} {\bibinfo {author} {\bibfnamefont {G.}~\bibnamefont
  {Sent{\'i}s}}, \bibinfo {author} {\bibfnamefont {M.}~\bibnamefont
  {Gu{\c{T}}{\u{a}}}}, \ and\ \bibinfo {author} {\bibfnamefont
  {G.}~\bibnamefont {Adesso}},\ }\bibfield  {title} {\enquote {\bibinfo {title}
  {Quantum learning of coherent states},}\ }\href {\doibase
  10.1140/epjqt/s40507-015-0030-4} {\bibfield  {journal} {\bibinfo  {journal}
  {EPJ Quantum Technology}\ }\textbf {\bibinfo {volume} {2}},\ \bibinfo {pages}
  {17} (\bibinfo {year} {2015})}\BibitemShut {NoStop}%
\bibitem [{\citenamefont {Banchi}\ \emph {et~al.}(2016)\citenamefont {Banchi},
  \citenamefont {Pancotti},\ and\ \citenamefont {Bose}}]{Banchi16}%
  \BibitemOpen
  \bibfield  {author} {\bibinfo {author} {\bibfnamefont {L.}~\bibnamefont
  {Banchi}}, \bibinfo {author} {\bibfnamefont {N.}~\bibnamefont {Pancotti}}, \
  and\ \bibinfo {author} {\bibfnamefont {S.}~\bibnamefont {Bose}},\ }\bibfield
  {title} {\enquote {\bibinfo {title} {Quantum gate learning in qubit networks:
  Toffoli gate without time-dependent control},}\ }\href@noop {} {\bibfield
  {journal} {\bibinfo  {journal} {Npj Quantum Information}\ }\textbf {\bibinfo
  {volume} {2}},\ \bibinfo {pages} {16019 EP --} (\bibinfo {year}
  {2016})}\BibitemShut {NoStop}%
\bibitem [{\citenamefont {Palittapongarnpim}\ \emph {et~al.}(2016)\citenamefont
  {Palittapongarnpim}, \citenamefont {Wittek}, \citenamefont {Zahedinejad},
  \citenamefont {Vedaie},\ and\ \citenamefont {Sanders}}]{Palittapongarnpim16}%
  \BibitemOpen
  \bibfield  {author} {\bibinfo {author} {\bibfnamefont {P.}~\bibnamefont
  {Palittapongarnpim}}, \bibinfo {author} {\bibfnamefont {P.}~\bibnamefont
  {Wittek}}, \bibinfo {author} {\bibfnamefont {E.}~\bibnamefont {Zahedinejad}},
  \bibinfo {author} {\bibfnamefont {S}~\bibnamefont {Vedaie}}, \ and\ \bibinfo
  {author} {\bibfnamefont {B.~C.}\ \bibnamefont {Sanders}},\ }\bibfield
  {title} {\enquote {\bibinfo {title} {Learning in quantum control:
  high-dimensional global optimization for noisy quantum dynamics},}\
  }\href@noop {} {\  (\bibinfo {year} {2016})}\BibitemShut {NoStop}%
\bibitem [{\citenamefont {Feynman}(1986)}]{Feynman86}%
  \BibitemOpen
  \bibfield  {author} {\bibinfo {author} {\bibfnamefont {R.~P.}\ \bibnamefont
  {Feynman}},\ }\bibfield  {title} {\enquote {\bibinfo {title} {Quantum
  mechanical computers},}\ }\href@noop {} {\bibfield  {journal} {\bibinfo
  {journal} {Found. Physics}\ }\textbf {\bibinfo {volume} {16}},\ \bibinfo
  {pages} {507–531} (\bibinfo {year} {1986})}\BibitemShut {NoStop}%
\bibitem [{\citenamefont {Muthukrishnan}(1999)}]{Muthukrishnan99}%
  \BibitemOpen
  \bibfield  {author} {\bibinfo {author} {\bibfnamefont {A.}~\bibnamefont
  {Muthukrishnan}},\ }\href
  {http://www.optics.rochester.edu/~stroud/presentations/muthukrishnan991/LogicGates.pdf}
  {\enquote {\bibinfo {title} {Classical and quantum logic gates: An
  introduction to quantum computing},}\ } (\bibinfo {year} {1999}),\ \bibinfo
  {note} {seminar notes, unpublished}\BibitemShut {NoStop}%
\bibitem [{\citenamefont {Curtis}\ and\ \citenamefont
  {Reiner}(1962)}]{CurtisReiner62}%
  \BibitemOpen
  \bibfield  {author} {\bibinfo {author} {\bibfnamefont {C.~W.}\ \bibnamefont
  {Curtis}}\ and\ \bibinfo {author} {\bibfnamefont {I.}~\bibnamefont
  {Reiner}},\ }\href@noop {} {\emph {\bibinfo {title} {Representation Theory of
  Finite Groups and Associative Algebras}}}\ (\bibinfo  {publisher} {AMS
  Chelsea Publishing},\ \bibinfo {address} {Providence, RI},\ \bibinfo {year}
  {1962})\BibitemShut {NoStop}%
\bibitem [{\citenamefont {Bartlett}\ and\ \citenamefont
  {Downs}(1992)}]{BartlettDowns92}%
  \BibitemOpen
  \bibfield  {author} {\bibinfo {author} {\bibfnamefont {P.~L.}\ \bibnamefont
  {Bartlett}}\ and\ \bibinfo {author} {\bibfnamefont {T.}~\bibnamefont
  {Downs}},\ }\bibfield  {title} {\enquote {\bibinfo {title} {Using random
  weights to train multilayer networks of hard-limiting units},}\ }\href@noop
  {} {\bibfield  {journal} {\bibinfo  {journal} {IEEE Transactions on Neural
  Networks}\ }\textbf {\bibinfo {volume} {3}},\ \bibinfo {pages} {202--210}
  (\bibinfo {year} {1992})}\BibitemShut {NoStop}%
\bibitem [{\citenamefont {Downs}\ and\ \citenamefont
  {Gaynier}(1995)}]{DownsGaynier95}%
  \BibitemOpen
  \bibfield  {author} {\bibinfo {author} {\bibfnamefont {T.}~\bibnamefont
  {Downs}}\ and\ \bibinfo {author} {\bibfnamefont {R.~J.}\ \bibnamefont
  {Gaynier}},\ }\bibfield  {title} {\enquote {\bibinfo {title} {The use of
  random weights for the training of multilayer networks of neurons with
  heaviside characteristics},}\ }\href@noop {} {\bibfield  {journal} {\bibinfo
  {journal} {Mathl. Comput. Modelling}\ }\textbf {\bibinfo {volume} {22}},\
  \bibinfo {pages} {53--61} (\bibinfo {year} {1995})}\BibitemShut {NoStop}%
\bibitem [{\citenamefont {Rowell}(2004)}]{Rowell04}%
  \BibitemOpen
  \bibfield  {author} {\bibinfo {author} {\bibfnamefont {D.}~\bibnamefont
  {Rowell}},\ }\href {http://web.mit.edu/2.151/www/Handouts/CayleyHamilton.pdf}
  {\enquote {\bibinfo {title} {Computing the matrix exponential the
  {C}ayley-{H}amilton method},}\ } (\bibinfo {year} {2004}),\ \bibinfo {note}
  {online lecture notes from MIT Dept. of Mech. Eng., unpublished}\BibitemShut
  {NoStop}%
\bibitem [{\citenamefont {Hedemann}(2013)}]{Hedemann13}%
  \BibitemOpen
  \bibfield  {author} {\bibinfo {author} {\bibfnamefont {S.~R.}\ \bibnamefont
  {Hedemann}},\ }\bibfield  {title} {\enquote {\bibinfo {title} {Hyperspherical
  parameterization of unitary matrices},}\ }\href@noop {} {\  (\bibinfo {year}
  {2013})},\ \Eprint {http://arxiv.org/abs/1303.5904} {arXiv:1303.5904
  [quant-ph]} \BibitemShut {NoStop}%
\bibitem [{\citenamefont {Wilde}(2013)}]{Wilde13}%
  \BibitemOpen
  \bibfield  {author} {\bibinfo {author} {\bibfnamefont {M.~M.}\ \bibnamefont
  {Wilde}},\ }\href@noop {} {\emph {\bibinfo {title} {Quantum Information
  Theory}}}\ (\bibinfo  {publisher} {Cambridge University Press},\ \bibinfo
  {address} {Cambridge},\ \bibinfo {year} {2013})\BibitemShut {NoStop}%
\bibitem [{\citenamefont {Rudolph}(2016)}]{silicon}%
  \BibitemOpen
  \bibfield  {author} {\bibinfo {author} {\bibfnamefont {T.}~\bibnamefont
  {Rudolph}},\ }\bibfield  {title} {\enquote {\bibinfo {title} {Why {I} am
  optimistic about the silicon-photonic route to quantum computing},}\
  }\href@noop {} {\  (\bibinfo {year} {2016})},\ \Eprint
  {http://arxiv.org/abs/1607.08535} {arXiv:1607.08535 [quant-ph]} \BibitemShut
  {NoStop}%
\bibitem [{\citenamefont {Rojas}(1996)}]{Rojas}%
  \BibitemOpen
  \bibfield  {author} {\bibinfo {author} {\bibfnamefont {R.}~\bibnamefont
  {Rojas}},\ }\href@noop {} {\emph {\bibinfo {title} {Neural Networks}}}\
  (\bibinfo  {publisher} {Springer},\ \bibinfo {address} {Berlin},\ \bibinfo
  {year} {1996})\BibitemShut {NoStop}%
\bibitem [{\citenamefont {Cerf}\ \emph {et~al.}(1998)\citenamefont {Cerf},
  \citenamefont {Adami},\ and\ \citenamefont {Kwiat}}]{CAK}%
  \BibitemOpen
  \bibfield  {author} {\bibinfo {author} {\bibfnamefont {N.~J.}\ \bibnamefont
  {Cerf}}, \bibinfo {author} {\bibfnamefont {C.}~\bibnamefont {Adami}}, \ and\
  \bibinfo {author} {\bibfnamefont {P.~G.}\ \bibnamefont {Kwiat}},\ }\bibfield
  {title} {\enquote {\bibinfo {title} {Optical simulation of quantum logic},}\
  }\href {\doibase 10.1103/PhysRevA.57.R1477} {\bibfield  {journal} {\bibinfo
  {journal} {Phys. Rev. A}\ }\textbf {\bibinfo {volume} {57}},\ \bibinfo
  {pages} {R1477--R1480} (\bibinfo {year} {1998})}\BibitemShut {NoStop}%
\bibitem [{\citenamefont {Reck}\ \emph {et~al.}(1994)\citenamefont {Reck},
  \citenamefont {Zeilinger}, \citenamefont {Bernstein},\ and\ \citenamefont
  {Bertani}}]{Reck}%
  \BibitemOpen
  \bibfield  {author} {\bibinfo {author} {\bibfnamefont {M.}~\bibnamefont
  {Reck}}, \bibinfo {author} {\bibfnamefont {A.}~\bibnamefont {Zeilinger}},
  \bibinfo {author} {\bibfnamefont {H.~J.}\ \bibnamefont {Bernstein}}, \ and\
  \bibinfo {author} {\bibfnamefont {P.}~\bibnamefont {Bertani}},\ }\bibfield
  {title} {\enquote {\bibinfo {title} {Experimental realization of any discrete
  unitary operator},}\ }\href {\doibase 10.1103/PhysRevLett.73.58} {\bibfield
  {journal} {\bibinfo  {journal} {Phys. Rev. Lett.}\ }\textbf {\bibinfo
  {volume} {73}},\ \bibinfo {pages} {58--61} (\bibinfo {year}
  {1994})}\BibitemShut {NoStop}%
\bibitem [{\citenamefont {Clements}\ \emph {et~al.}(2016)\citenamefont
  {Clements}, \citenamefont {Humphreys}, \citenamefont {Metcalf}, \citenamefont
  {Kolthammer},\ and\ \citenamefont {Walmsley}}]{Clements}%
  \BibitemOpen
  \bibfield  {author} {\bibinfo {author} {\bibfnamefont {W.~R.}\ \bibnamefont
  {Clements}}, \bibinfo {author} {\bibfnamefont {P.~C.}\ \bibnamefont
  {Humphreys}}, \bibinfo {author} {\bibfnamefont {B.~J.}\ \bibnamefont
  {Metcalf}}, \bibinfo {author} {\bibfnamefont {W.~S.}\ \bibnamefont
  {Kolthammer}}, \ and\ \bibinfo {author} {\bibfnamefont {I.~A.}\ \bibnamefont
  {Walmsley}},\ }\bibfield  {title} {\enquote {\bibinfo {title} {An optimal
  design for universal multiport interferometers},}\ }\href@noop {} {\
  (\bibinfo {year} {2016})},\ \Eprint {http://arxiv.org/abs/1603.08788}
  {arXiv:1603.08788 [physics.optics]} \BibitemShut {NoStop}%
\bibitem [{\citenamefont {Knill}\ \emph {et~al.}(2001)\citenamefont {Knill},
  \citenamefont {Laflamme},\ and\ \citenamefont {Milburn}}]{klm}%
  \BibitemOpen
  \bibfield  {author} {\bibinfo {author} {\bibfnamefont {E.}~\bibnamefont
  {Knill}}, \bibinfo {author} {\bibfnamefont {R.}~\bibnamefont {Laflamme}}, \
  and\ \bibinfo {author} {\bibfnamefont {G.~J.}\ \bibnamefont {Milburn}},\
  }\bibfield  {title} {\enquote {\bibinfo {title} {A scheme for efficient
  quantum computation with linear optics},}\ }\href {\doibase 10.1038/35051009}
  {\bibfield  {journal} {\bibinfo  {journal} {Nature}\ }\textbf {\bibinfo
  {volume} {409}},\ \bibinfo {pages} {46--52} (\bibinfo {year}
  {2001})}\BibitemShut {NoStop}%
\bibitem [{\citenamefont {Humphreys}\ \emph {et~al.}(2014)\citenamefont
  {Humphreys}, \citenamefont {Metcalf}, \citenamefont {Spring}, \citenamefont
  {Moore}, \citenamefont {Salter}, \citenamefont {Booth}, \citenamefont
  {Kolthammer},\ and\ \citenamefont {Walmsley}}]{strain}%
  \BibitemOpen
  \bibfield  {author} {\bibinfo {author} {\bibfnamefont {P.~C.}\ \bibnamefont
  {Humphreys}}, \bibinfo {author} {\bibfnamefont {B.~J.}\ \bibnamefont
  {Metcalf}}, \bibinfo {author} {\bibfnamefont {J.~B.}\ \bibnamefont {Spring}},
  \bibinfo {author} {\bibfnamefont {M.}~\bibnamefont {Moore}}, \bibinfo
  {author} {\bibfnamefont {P.~S.}\ \bibnamefont {Salter}}, \bibinfo {author}
  {\bibfnamefont {M.~J.}\ \bibnamefont {Booth}}, \bibinfo {author}
  {\bibfnamefont {W.~S.}\ \bibnamefont {Kolthammer}}, \ and\ \bibinfo {author}
  {\bibfnamefont {I.~A.}\ \bibnamefont {Walmsley}},\ }\bibfield  {title}
  {\enquote {\bibinfo {title} {Strain-optic active control for quantum
  integrated photonics},}\ }\href {\doibase 10.1364/OE.22.021719} {\bibfield
  {journal} {\bibinfo  {journal} {Opt. Express}\ }\textbf {\bibinfo {volume}
  {22}},\ \bibinfo {pages} {21719--21726} (\bibinfo {year} {2014})}\BibitemShut
  {NoStop}%
\bibitem [{\citenamefont {Sansoni}\ \emph {et~al.}(2010)\citenamefont
  {Sansoni}, \citenamefont {Sciarrino}, \citenamefont {Vallone}, \citenamefont
  {Mataloni}, \citenamefont {Crespi}, \citenamefont {Ramponi},\ and\
  \citenamefont {Osellame}}]{polarisationstates}%
  \BibitemOpen
  \bibfield  {author} {\bibinfo {author} {\bibfnamefont {L.}~\bibnamefont
  {Sansoni}}, \bibinfo {author} {\bibfnamefont {F.}~\bibnamefont {Sciarrino}},
  \bibinfo {author} {\bibfnamefont {G.}~\bibnamefont {Vallone}}, \bibinfo
  {author} {\bibfnamefont {P.}~\bibnamefont {Mataloni}}, \bibinfo {author}
  {\bibfnamefont {A.}~\bibnamefont {Crespi}}, \bibinfo {author} {\bibfnamefont
  {R.}~\bibnamefont {Ramponi}}, \ and\ \bibinfo {author} {\bibfnamefont
  {R.}~\bibnamefont {Osellame}},\ }\bibfield  {title} {\enquote {\bibinfo
  {title} {Polarization entangled state measurement on a chip},}\ }\href
  {\doibase 10.1103/PhysRevLett.105.200503} {\bibfield  {journal} {\bibinfo
  {journal} {Phys. Rev. Lett.}\ }\textbf {\bibinfo {volume} {105}},\ \bibinfo
  {pages} {200503} (\bibinfo {year} {2010})}\BibitemShut {NoStop}%
\bibitem [{\citenamefont {Bonneau}\ \emph {et~al.}(2012)\citenamefont
  {Bonneau}, \citenamefont {Lobino}, \citenamefont {Jiang}, \citenamefont
  {Natarajan}, \citenamefont {Tanner}, \citenamefont {Hadfield}, \citenamefont
  {Dorenbos}, \citenamefont {Zwiller}, \citenamefont {Thompson},\ and\
  \citenamefont {O'Brien}}]{polarisationshift}%
  \BibitemOpen
  \bibfield  {author} {\bibinfo {author} {\bibfnamefont {D.}~\bibnamefont
  {Bonneau}}, \bibinfo {author} {\bibfnamefont {M.}~\bibnamefont {Lobino}},
  \bibinfo {author} {\bibfnamefont {P.}~\bibnamefont {Jiang}}, \bibinfo
  {author} {\bibfnamefont {C.~M.}\ \bibnamefont {Natarajan}}, \bibinfo {author}
  {\bibfnamefont {M.~G.}\ \bibnamefont {Tanner}}, \bibinfo {author}
  {\bibfnamefont {R.~H.}\ \bibnamefont {Hadfield}}, \bibinfo {author}
  {\bibfnamefont {S.~N.}\ \bibnamefont {Dorenbos}}, \bibinfo {author}
  {\bibfnamefont {V.}~\bibnamefont {Zwiller}}, \bibinfo {author} {\bibfnamefont
  {M.~G.}\ \bibnamefont {Thompson}}, \ and\ \bibinfo {author} {\bibfnamefont
  {J.~L.}\ \bibnamefont {O'Brien}},\ }\bibfield  {title} {\enquote {\bibinfo
  {title} {Fast path and polarization manipulation of telecom wavelength single
  photons in lithium niobate waveguide devices},}\ }\href {\doibase
  10.1103/PhysRevLett.108.053601} {\bibfield  {journal} {\bibinfo  {journal}
  {Phys. Rev. Lett.}\ }\textbf {\bibinfo {volume} {108}},\ \bibinfo {pages}
  {053601} (\bibinfo {year} {2012})}\BibitemShut {NoStop}%
\end{thebibliography}%

\end{document}